\documentclass[aps,prd,onecolumn,groupedaddress,showpacs,nofootinbib,amssymb]{revtex4}
\usepackage[dvips]{graphicx}
\usepackage{amssymb}
\usepackage{amsmath}
\usepackage{graphicx,color}
\usepackage{amsfonts}
\usepackage{bm}
\usepackage{cancel}
\usepackage{comment}

\newcommand{\be}{\begin{equation}}
\newcommand{\ee}{\end{equation}}
\newcommand{\bea}{\begin{eqnarray}}
\newcommand{\eea}{\end{eqnarray}}
\newcommand{\beaa}{\begin{eqnarray*}}
\newcommand{\eeaa}{\end{eqnarray*}}

\newcommand{\nn}{\nonumber \\}
\newcommand{\e}{\mathrm{e}}

\allowdisplaybreaks[4]
\begin{document}

\title{Propagation of Gravitational Waves in Chern-Simons Axion Einstein Gravity}
\author{Shin'ichi~Nojiri,$^{1,2}$\,\thanks{nojiri@gravity.phys.nagoya-u.ac.jp}
S.~D.~Odintsov,$^{3,4,5}$\,\thanks{odintsov@ieec.uab.es}
V.~K.~Oikonomou,$^{6,7,8}$\,\thanks{v.k.oikonomou1979@gmail.com}
Arkady A. Popov,$^{9}$\,\thanks{arkady_popov@mail.ru}}
\affiliation{ $^{1)}$ Department of Physics, Nagoya University,
Nagoya 464-8602, Japan \\
$^{2)}$ Kobayashi-Maskawa Institute for the Origin of Particles
and the Universe, Nagoya University, Nagoya 464-8602, Japan \\
$^{3)}$ ICREA, Passeig Luis Companys, 23, 08010 Barcelona, Spain\\
$^{4)}$ Institute of Space Sciences (IEEC-CSIC) C. Can Magrans s/n,
08193 Barcelona, Spain\\
$^{5)}$ Tomsk State Pedagogical University, 634061 Tomsk, Russia\\
$^{6)}$ Department of Physics, Aristotle University of
Thessaloniki, Thessaloniki 54124, Greece\\
$^{7)}$ International Laboratory for Theoretical Cosmology, Tomsk
State University of Control Systems
and Radioelectronics (TUSUR), 634050 Tomsk, Russia\\
$^{8)}$ Theoretical Astrophysics, IAAT, University of T\"{u}bingen, Germany
\\
$^{9)}$ N. I. Lobachevsky Institute of Mathematics and Mechanics, Kazan Federal University,
420008, Kremlevskaya street 18, Kazan, Russia}

\tolerance=5000

\begin{abstract}
In this paper we shall investigate the propagation of gravitational waves in a flat Friedman-Robertson-Walker background, in the context of a string motivated corrected Einstein gravity. Particularly, we shall consider a misalignment axion Einstein gravity in the presence of a string originating Chern-Simons coupling of the axion field to the Chern-Pontryagin density in four dimensions. We shall focus our study on the propagation of the gravitational waves, and we shall investigate whether there exists any difference in the propagation of the polarization states of the gravitational waves. As we demonstrate, the dispersion relations are different in the Right-handed mode and the Left-handed mode. Finally, we compare the propagation of the axion Chern-Simons Einstein theory with that of standard $F(R)$ gravity.
\end{abstract}


\maketitle

\section{Introduction}

String theory is one solid theoretical framework that may describe in a consistent way
the ultraviolet completion of classical gravity and of the Standard Model of particle physics.
In some cases, several low-energy string theory effects may have their impact in
the classical gravitational phenomena.
One such example is offered by axion like particles \cite{Marsh:2015xka,
Marsh:2017yvc,Odintsov:2019mlf,Nojiri:2019riz,Odintsov:2019evb,Cicoli:2019ulk,
Fukunaga:2019unq,Caputo:2019joi,Sakharov:1994id,Sakharov:1996xg,Khlopov:1999tm},
and specifically from the misalignment axions \cite{Marsh:2015xka}.
For these axions, there exist a primordial $U(1)$ symmetry which is unbroken quite earlier
from the inflationary era, but it is broken during the inflationary era.
The cosmological implications of the axions have quite appealing features, and in view of
the growing research stream related with axion experiments and observations,
\cite{Du:2018uak,Henning:2018ogd,Ouellet:2018beu,Safdi:2018oeu,Rozner:2019gba,
Avignone:2018zpw,Caputo:2018vmy,Caputo:2018ljp,Balakin:2009rg,Balakin:2012up,
Balakin:2014oya},
see also \cite{Lawson:2019brd}, axion physics has become quite timely.
Actually, the axion may be one of the last realistic Weakly Interacting Massive Particles
(WIMP) \cite{Oikonomou:2006mh}, since a longstanding number of experiments focusing
on large WIMP masses, ended up with no result.
Thus, unless supersymmetry is discovered in the Large Hadron Collider, the axion seems to
be the last realistic particle dark matter candidate.

Very closely related to axion gravity, the so-called Chern-Simons term
\cite{Nishizawa:2018srh,Wagle:2018tyk,Yagi:2012vf,Yagi:2012ya,Molina:2010fb,
Izaurieta:2009hz,Sopuerta:2009iy,Konno:2009kg,Smith:2007jm,Matschull:1999he,
Haghani:2017yjk,Satoh:2007gn,Satoh:2008ck,Yoshida:2017cjl,
Alexander:2007kv,Yunes:2010yf,Alexander:2017jmt} coupled with a function
depending on the axion field, may be a realistic string theory originating correction, that
may have a direct impact on low-energy gravitational phenomena and even in the
inflationary era itself.
The Chern-Simons term coupled to the axion has the form $U(\phi)\tilde{R}R$, and it just
the Chern-Pontryagin density.
This string motivated term can have quite interesting effects on inflationary physics, as it
was demonstrated firstly in Refs.~\cite{Hwang:2005hb,Choi:1999zy}, see also
Refs.~\cite{Odintsov:2019mlf,Nojiri:2019riz,Odintsov:2019evb} for some recent modified
gravity applications.
As it was shown in \cite{Hwang:2005hb}, the Chern-Simons term can affect directly the
tensor modes of the primordial perturbations, and this can have a measurable effect on the
tensor-to-scalar ratio. This feature was also shown to occur in the context of
the Chern-Simons $F(R)$ gravity \cite{Odintsov:2019mlf}.
Also in Ref.~\cite{Choi:1999zy}, it was demonstrated that the primordial gravitational
waves have two polarizations that propagate in a different way.
This phenomenon is similar and is related to chiral gravitational waves, which are also a
very timely subject of current research
\cite{Inomata:2018rin,Kamionkowski:1997av,Pritchard:2004qp,Lyth:2005jf,Alexander:2007qe,Alexander:2004us,Cai:2016ihp,Moretti:2019yhs}.

Motivated by the above, in this work we shall investigate the propagation of gravitational
waves, in the presence of the Chern-Simons axion term in the context of Einstein gravity.
Our aim is to discover whether the dispersion relations are different for the two polarization
modes, that is, the Right-handed mode and the Left-handed mode,
which are given by the linear combinations of the $+$-mode and $\times$-mode,
focusing on the present time era, since it is known that after horizon crossing, the
amplitudes of the gravitational waves are conserved in the large scale limit
\cite{Choi:1999zy}. Our findings are quite interesting, since we demonstrate that during a
late-time nearly de Sitter era, if the axion coupling term is nearly constant, there exists
a non-trivial mixing between the polarization states, and the two polarizations have
different dispersion relations. In addition, just for comparison reasons, we discuss the
propagation of gravitational waves in the context of $F(R)$ gravity, which is the most
representative theory of modified gravity \cite{Nojiri:2017ncd,Nojiri:2010wj,Nojiri:2006ri,
Capozziello:2011et,Capozziello:2010zz,delaCruzDombriz:2012xy,Olmo:2011uz}, and we
compare the results to the Chern-Simons axion Einstein theory.

This paper is organized as follows: In section II, we discuss several geometric features of
the Einstein Chern-Simons gravity, which we shall extensively use
in the next sections.
In section III, we present the essential features of the Einstein Chern-Simons theory,
and we study in detail the propagation of gravitational waves in a flat
Friedmann-Robertson-Walker background.
Finally, in section IV, we discuss in detail the propagation of gravitational waves in
$F(R)$ gravity, and we compare the results of $F(R)$ gravity with the results we found
corresponding to the Einstein Chern-Simons gravity.

\section{Geometric Background and Formalism}

Before getting to the core of our study, we shall briefly elaborate on the geometric features of the theory at hand. The curvature tensors and the connections are,
\begin{align}
\label{RI}
R =&g^{\mu\nu}R_{\mu\nu} \, , \quad
R_{\mu\nu} = R^\lambda_{\ \mu\lambda\nu} \, , \quad
R^\lambda_{\ \mu\rho\nu}= -\Gamma^\lambda_{\mu\rho,\nu}
+ \Gamma^\lambda_{\mu\nu,\rho} - \Gamma^\eta_{\mu\rho}\Gamma^\lambda_{\nu\eta}
+ \Gamma^\eta_{\mu\nu}\Gamma^\lambda_{\rho\eta} \, , \nn
\Gamma^\eta_{\mu\lambda} =& \frac{1}{2}g^{\eta\nu}\left(
g_{\mu\nu,\lambda} + g_{\lambda\nu,\mu} - g_{\mu\lambda,\nu} \right)\, .
\end{align}
By varying the metric tensor $g_{\mu \nu}$ as follows,
\begin{equation}
\label{Q3}
g_{\mu\nu}\to g_{\mu\nu} + \delta g_{\mu\nu}\, ,
\end{equation}
we obtain the variation of the following tensors, which we shall need and use extensively
in the following sections,
\begin{align}
\label{Q4}
\delta\Gamma^\kappa_{\mu\nu} =&\frac{1}{2}g^{\kappa\lambda}\left(
\nabla_\mu \delta g_{\nu\lambda} + \nabla_\nu \delta g_{\mu\lambda}
 - \nabla_\lambda \delta g_{\mu\nu} \right)\, ,\nn
\delta R^\mu_{\ \nu\lambda\sigma}
=& \nabla_\lambda \delta\Gamma^\mu_{\sigma\nu}
 - \nabla_\sigma \delta \Gamma^\mu_{\lambda\nu}\ ,\nn
\delta R_{\mu\nu\lambda\sigma}
=&\frac{1}{2}\left[\nabla_\lambda \nabla_\nu \delta g_{\sigma\mu}
 - \nabla_\lambda \nabla_\mu \delta g_{\sigma\nu}
 - \nabla_\sigma \nabla_\nu \delta g_{\lambda\mu}
+ \nabla_\sigma \nabla_\mu \delta g_{\lambda\nu}
+ \delta g_{\mu\rho} R^\rho_{\ \nu\lambda\sigma}
 - \delta g_{\nu\rho} R^\rho_{\ \mu\lambda\sigma} \right] \, ,\nn
\delta R_{\mu\nu} =& \frac{1}{2}\left[\nabla^\rho\left(\nabla_\mu \delta g_{\nu\rho}
+ \nabla_\nu \delta g_{\mu\rho}\right) - \nabla^2 \delta g_{\mu\nu}
 - \nabla_\mu \nabla_\nu \left(g^{\rho\lambda}\delta g_{\rho\lambda}\right)\right] \nn
=& \frac{1}{2}\left[\nabla_\mu\nabla^\rho \delta g_{\nu\rho}
+ \nabla_\nu \nabla^\rho \delta g_{\mu\rho} - \nabla^2 \delta g_{\mu\nu}
 - \nabla_\mu \nabla_\nu \left(g^{\rho\lambda}\delta g_{\rho\lambda}\right) 
 - 2R^{\lambda\ \rho}_{\ \nu\ \mu}\delta g_{\lambda\rho}
+ R^\rho_{\ \mu}\delta g_{\rho\nu} + R^\rho_{\ \mu}\delta g_{\rho\nu} \right]\, ,\nn
\delta R =& -\delta g_{\mu\nu} R^{\mu\nu} + \nabla^\mu \nabla^\nu \delta g_{\mu\nu}
 - \nabla^2 \left(g^{\mu\nu}\delta g_{\mu\nu}\right)\, , \nn
\left[\nabla_\mu,\nabla_\nu\right]\delta g_{\lambda\sigma}
=& - \delta g_{\rho\sigma} R^\rho_{\ \lambda\mu\nu}
 - \delta g_{\lambda\rho} R^\rho_{\ \sigma\mu\nu} \, .
\end{align}
Consider now the anti-symmetric tensor $A_{\mu\nu\rho\sigma}$,
and with it we construct the following action,
\begin{equation}
\label{CS1}
S_A = \int dx^\mu dx^\nu dx^\rho dx^\sigma A_{\mu\nu\rho\sigma}\, ,
\end{equation}
which is invariant under the general coordinate transformation.
If we define the totally antisymmetric Levi-Civita symbols
$\epsilon_{\mu\nu\rho\sigma}$ and $\epsilon^{\mu\nu\rho\sigma}$ as follows,
\begin{equation}
\label{CS2}
\epsilon_{0123} = - \epsilon^{0123} = 1 \, ,
\end{equation}
we have the following,
\begin{align}
\label{CS3}
dx^\mu dx^\nu dx^\rho dx^\sigma A_{\mu\nu\rho\sigma}
=& - \frac{1}{4!} dx^\mu dx^\nu dx^\rho dx^\sigma \epsilon_{\mu\nu\rho\sigma}
A_{\mu'\nu'\rho'\sigma'} \epsilon^{\mu'\nu'\rho'\sigma'} \nn
=& d^4 x \sqrt{-g} \frac{1}{\sqrt{-g}}
A_{\mu'\nu'\rho'\sigma'} \epsilon^{\mu'\nu'\rho'\sigma'} \, .
\end{align}
Then if we define $\tilde\epsilon^{\mu\nu\rho\sigma}$ by,
\begin{equation}
\label{CS4}
\tilde\epsilon^{\mu\nu\rho\sigma} \equiv
\frac{1}{\sqrt{-g}}\epsilon^{\mu\nu\rho\sigma} \, ,
\end{equation}
we can regard $\tilde\epsilon^{\mu\nu\rho\sigma}$ as a tensor. We can also define,
\begin{equation}
\label{CS4B}
\epsilon_{\mu\nu\rho\sigma}
= \eta_{\mu\mu'} \eta_{\nu\nu'} \eta_{\rho\rho'} \eta_{\sigma\sigma'}
\epsilon^{\mu'\nu'\rho'\sigma'} \, ,
\end{equation}
and in addition,
\begin{equation}
\label{CS4C}
\tilde \epsilon_{\mu\nu\rho\sigma}
\equiv g_{\mu\mu'} g_{\nu\nu'} g_{\rho\rho'} g_{\sigma\sigma'}
\tilde\epsilon^{\mu'\nu'\rho'\sigma'}
= \sqrt{-g} \epsilon_{\mu\nu\rho\sigma} \, .
\end{equation}
We should note that,
\begin{equation}
\label{CS4D}
\nabla_\sigma \tilde\epsilon^{\zeta\eta\rho\xi} =0 \, .
\end{equation}
due to the fact that,
\begin{align}
\label{CS4E}
& \tilde\epsilon_{\zeta\eta\rho\xi}
\nabla_\sigma \tilde\epsilon^{\zeta\eta\rho\xi} \nn
= & \tilde\epsilon_{\zeta\eta\rho\xi}
\left( \partial _\sigma \tilde\epsilon^{\zeta\eta\rho\xi}
+ \Gamma^\zeta_{\sigma\tau} \tilde\epsilon^{\tau\eta\rho\xi}
+ \Gamma^\eta_{\sigma\tau} \tilde\epsilon^{\zeta\tau\rho\xi}
+ \Gamma^\rho_{\sigma\tau} \tilde\epsilon^{\zeta\eta\tau\xi}
+ \Gamma^\xi_{\sigma\tau} \tilde\epsilon^{\zeta\eta\rho\tau} \right) \nn
= & - \frac{4!}{2} g^{\mu\nu} \partial _\sigma g_{\mu\nu}
+ 3! \left( \Gamma^\zeta_{\sigma\zeta}
+ \Gamma^\eta_{\sigma\eta}
+ \Gamma^\rho_{\sigma\rho}
+ \Gamma^\xi_{\sigma\xi} \right) \nn
= & \frac{4!}{2} \left( - g^{\mu\nu} \partial _\sigma g_{\mu\nu}
+ g^{\tau\mu} \left( \partial_\sigma g_{\mu\tau}
+ \partial_\tau g_{\sigma\mu} - \partial_\mu g_{\sigma\tau} \right)
\right) \nn
= & 0\, .
\end{align}
We should note that there is only one non-vanishing component
in $\tilde\epsilon^{\zeta\eta\rho\xi}$ and therefore if
$\tilde\epsilon_{\zeta\eta\rho\xi}
\nabla_\sigma \tilde\epsilon^{\zeta\eta\rho\xi}$ vanishes, the quantity
$\nabla_\sigma \tilde\epsilon^{\zeta\eta\rho\xi}$ also vanishes. The above identities
and quantities defined, shall be extensively used in the following section.

\section{The Chern-Simons Corrected Axion Model}

We shall consider a Chern-Simons corrected misalignment axion in the context of Einstein gravity, so the action has the following form,
\begin{equation}
\label{CS5}
S = \frac{1}{2\kappa^2} \int d^4 x \sqrt{-g} \left[
R - \frac{\omega(\phi)}{2}\partial_\mu \phi \partial^\mu \phi - V(\phi)
+ U(\phi) \tilde\epsilon^{\mu\nu\rho\sigma} R^\tau_{\ \lambda\mu\nu}
R^\lambda_{\ \tau\rho\sigma} \right] \, .
\end{equation}
In most cases, the misalignment axion is taken to be a canonical scalar field, but we
assume that it has a generalized kinetic term, for generality of the argument.
Due to the fact that,
\begin{equation}
\label{CS6}
\delta \left( \sqrt{-g} U(\phi) \tilde\epsilon^{\mu\nu\rho\sigma}
R^\tau_{\ \lambda\mu\nu}
R^\lambda_{\ \tau\rho\sigma} \right)
= 2 \sqrt{-g} U(\phi) \left[ \tilde\epsilon^{\zeta\eta\rho\mu}
R^{\tau\nu}_{\ \ \zeta\eta} + \tilde\epsilon^{\zeta\eta\rho\nu}
R^{\tau\mu}_{\ \ \zeta\eta} \right] \nabla_\rho \nabla_\tau
\delta g_{\mu\nu}
\, ,
\end{equation}
by varying Eq.~(\ref{Q3}), the action (\ref{CS5}) gives the following equations of motion,
\begin{align}
\label{CS7}
0 =& \frac{1}{2} \left\{ R - \frac{\omega(\phi)}{2}\partial_\mu \phi \partial^\mu \phi
 - V(\phi) \right\} g_{\mu\nu}
 - R_{\mu\nu} + \frac{\omega(\phi)}{2}\partial_\mu \phi \partial_\nu \phi \nn
& + 2 \left( g_{\mu\xi} g_{\nu\sigma} + g_{\mu\sigma} g_{\nu\xi} \right)
\nabla_\tau \nabla_\rho \left( U(\phi) \tilde\epsilon^{\zeta\eta\rho\xi}
R^{\tau\sigma}_{\ \ \zeta\eta} \right) \, .
\end{align}

Let us in brief consider the solutions in a flat FRW spacetime, the line element of which is,
\begin{equation}
\label{FRWmetric}
ds^2 = - dt^2 + a(t)^2 \sum_{i=1,2,3} \left( dx^i \right)^2 \, ,
\end{equation}
and we shall assume that the scalar field $\phi$ depends solely on the cosmic time $t$.
Then, in view of the following relations that hold true for the flat FRW background,
\begin{align}
\label{E2}
& \Gamma^t_{ij}=a^2 H \delta_{ij}\, ,\quad \Gamma^i_{jt}=\Gamma^i_{tj}=H\delta^i_{\ j}\, ,
\quad \Gamma^i_{jk}=\tilde \Gamma^i_{jk}\, ,\nn
& R_{itjt}=-\left(\dot H + H^2\right)a^2\delta_{ij}\, ,\quad
R_{ijkl}= a^4 H^2 \left(\delta_{ik} \delta_{lj}
 - \delta_{il} \delta_{kj}\right)\, ,\nn
& R_{tt}=-3\left(\dot H + H^2\right)\, ,\quad
R_{ij}=a^2 \left(\dot H + 3H^2\right)\delta_{ij}\, ,\quad
R= 6\dot H + 12 H^2\, , \nn
& \mbox{other components}=0\, ,
\end{align}
we obtain the following two equations,
\begin{equation}
\label{CS8}
0 = - 3 H^2 + \frac{\omega(\phi)}{4}{\dot\phi}^2 + \frac{V(\phi)}{2} \, , \quad
0 = 2 \dot H + 3 H^2 + \frac{\omega(\phi)}{4}{\dot\phi}^2 - \frac{V(\phi)}{2} \, .
\end{equation}
We should note that the term containing the scalar coupling function to the Chern-Simons
term, namely, $U(\phi)$, does not contribute to the above equations in (\ref{CS8}).
This was also noted in Ref.~\cite{Hwang:2005hb} and actually, the Chern-Simons does not
affect the scalar perturbations at all, only the tensor perturbations.
We may choose $\phi$ to be the cosmological time $t$, that is, $\phi=t$.
Then the equations in (\ref{CS8}) have the following forms,
\begin{equation}
\label{CS9}
0 = - 3 H^2 + \frac{\omega(\phi)}{4} + \frac{V(\phi)}{2} \, , \quad
0 = 2 \dot H + 3 H^2 + \frac{\omega(\phi)}{4} - \frac{V(\phi)}{2}
\, ,
\end{equation}
which can be solved with respect to $\omega(\phi)$ and $V(\phi)$ as follows,
\begin{equation}
\label{CS10}
\omega(\phi) = - 4 \dot H \, , \quad V(\phi) = 2 \dot H + 6 H^2 \, .
\end{equation}
Then if we choose
\begin{equation}
\label{CS11}
\omega(\phi) = - 4 f'(\phi) \, , \quad
V(\phi) = 2 f'(\phi) + 6 f(\phi)^2 \, ,
\end{equation}
a solution of the equations in (\ref{CS8}) is given by $H=f(t)$ and $\phi=t$.
After these preliminary general features, which indicated that the Chern-Simons term does
not affect at all the equations of motion of the gravitational theory, we proceed to the study
of the gravity waves.
As we shall demonstrate in the next subsection, the gravitational waves are affected
significantly by the presence of the non-trivial Chern-Simons term.

\subsection{Gravitational Waves and Their Polarizations}

In this section we shall study the propagation of gravitational waves in the Chern-Simons axion Einstein gravity.
In order to study the propagation of the gravitational wave, we consider the perturbation of
Eq.~(\ref{CS7}), from the background whose metric is $g^{(0)}_{\mu\nu}$,
\begin{equation}
\label{CS14}
g_{\mu\nu} = g^{(0)}_{\mu\nu} + h_{\mu\nu} \, .
\end{equation}
Then the corresponding Einstein equations are,
\begin{align}
\label{CS15}
G_{\mu \nu}=& - \frac{1}{2}\Big(\nabla^{(0)}_\mu
\nabla^{(0)\, \rho} h_{\nu\rho}
+ \nabla^{(0)}_\nu \nabla^{(0)\, \rho} h_{\mu\rho} - \Box^{(0)} h_{\mu\nu}
 - \nabla^{(0)}_\mu \nabla^{(0)}_\nu \left(g^{(0)\, \rho\lambda} h_{\rho\lambda}\right)
\nn &
- 2R^{(0)\, \lambda\ \rho}_{\ \ \ \ \ \nu\ \mu} h_{\lambda\rho}
+ R^{(0)\, \rho}_{\ \ \ \ \ \mu} h_{\rho\nu}
+ R^{(0)\, \rho}_{\ \ \ \ \ \nu} h_{\rho\mu} \Big) \nn
& + \frac{1}{2} \left( R^{(0)}
 - \frac{\omega \left(\phi^{(0)} \right)}{2} g^{(0)\rho\sigma}\partial_\rho \phi^{(0)}
\partial_\sigma \phi^{(0)}  - V\left( \phi^{(0)} \right) \right) h_{\mu\nu} \nn
& + \frac{1}{2}g^{(0)}_{\mu\nu} \left( - h_{\rho\sigma}
\left( R^{(0)\, \rho\sigma} - \frac{\omega\left( \phi^{(0)} \right)}{2} \partial^\rho \phi^{(0)}
\partial^\sigma \phi^{(0)} \right)
+ \nabla^{(0)\, \rho} \nabla^{(0)\, \sigma} h_{\rho\sigma}
 - \Box^{(0)} \left(g^{(0)\, \rho\sigma} h_{\rho\sigma}\right) \right) \nn
& + 2 \left( h_{\mu\xi} g^{(0)}_{\nu\sigma} + h_{\mu\sigma} g^{(0)}_{\nu\xi}
+ g^{(0)}_{\mu\xi} h_{\nu\sigma} + g^{(0)}_{\mu\sigma} h_{\nu\xi} \right)
\tilde\epsilon^{(0)\, \zeta\eta\rho\xi}
\nabla^{(0)}_\tau \nabla^{(0)}_\rho \left( U \left(\phi^{(0)} \right)
R^{(0)\, \tau\sigma}_{\ \ \ \ \ \ \zeta\eta} \right) \nn
& + 2 \left( g^{(0)}_{\mu\xi} g^{(0)}_{\nu\sigma}
+ g^{(0)}_{\mu\sigma} g^{(0)}_{\nu\xi} \right) \left\{
 - \frac{1}{2} g^{(0)\, \alpha\beta} h_{\alpha\beta} \tilde\epsilon^{(0)\, \zeta\eta\rho\xi}
\nabla^{(0)}_\tau \nabla^{(0)}_\rho
\left( U \left(\phi^{(0)}\right) R^{(0)\, \tau\sigma}_{\ \ \ \ \ \ \zeta\eta} \right) \right. \nn
& + \tilde\epsilon^{(0)\, \zeta\eta\rho\xi} h_{\alpha\beta} \nabla^{(0)\, \alpha}
\nabla^{(0)}_\rho \left( U \left(\phi^{(0)} \right)
R^{(0)\, \sigma\beta}_{\ \ \ \ \ \ \zeta\eta} \right) \nn
& - \frac{1}{2} \tilde\epsilon^{(0)\, \zeta\eta\rho\xi} g^{(0)\, \tau\alpha}
\nabla^{(0)}_\tau \nabla^{(0)}_\rho
\left( 2 U \left(\phi^{(0)} \right) g^{(0)\, \sigma\beta}
\left( \nabla^{(0)}_\zeta \left( \nabla^{(0)}_\eta h_{\alpha\beta}
+ \nabla^{(0)}_\alpha h_{\eta\beta} - \nabla^{(0)}_\beta h_{\eta\alpha}
\right) \right) \right) \nn
& - \frac{1}{2} \tilde\epsilon^{(0)\, \zeta\eta\rho\xi}
g^{(0)\, \tau\alpha} \nabla^{(0)}_\tau \left( U \left(\phi^{(0)} \right) \left(
g^{(0)\, \sigma\beta} \left( \nabla^{(0)}_\rho h_{\beta\gamma}
+ \nabla^{(0)}_\gamma h_{\rho\beta} - \nabla^{(0)}_\beta h_{\rho\gamma} \right)
R^{(0)\, \gamma}_{\ \ \ \ \ \alpha\zeta\eta} \right. \right. \nn
& - g^{(0)\, \gamma\beta} \left( \nabla^{(0)}_\rho h_{\beta\alpha}
+ \nabla^{(0)}_\alpha h_{\rho\beta} - \nabla^{(0)}_\beta h_{\rho\alpha} \right)
R^{(0)\, \sigma}_{\ \ \ \ \ \gamma\zeta\eta} \nn
&  - g^{(0)\, \gamma\beta} \left( \nabla^{(0)}_\rho h_{\beta\zeta}
+ \nabla^{(0)}_\zeta h_{\rho\beta} - \nabla^{(0)}_\beta h_{\rho\zeta} \right)
R^{(0)\, \sigma}_{\ \ \ \ \ \alpha\gamma\eta} \nn
& \left. \left. - g^{(0)\, \gamma\beta} \left( \nabla^{(0)}_\rho h_{\beta\eta}
+ \nabla^{(0)}_\eta h_{\rho\beta} - \nabla^{(0)}_\beta h g_{\rho\eta} \right)
R^{(0)\, \sigma}_{\ \ \ \ \ \alpha\zeta\gamma} \right) \right) \nn
& - \frac{1}{2} \tilde\epsilon^{(0)\, \zeta\eta\rho\xi}
g^{(0) \tau\alpha} \left( - g^{(0)\ \beta\gamma} \left( \nabla^{(0)}_\tau h_{\gamma\rho}
+ \nabla^{(0)}_\rho h_{\tau\gamma} - \nabla^{(0)}_\gamma h_{\tau\rho} \right)
\nabla^{(0)}_\beta
\left( U \left(\phi^{(0)} \right) R^{(0)\, \sigma}_{\ \ \ \ \ \alpha\zeta\eta} \right)
\right. \nn
& + g^{(0)\, \sigma\gamma} \left( \nabla^{(0)}_\tau h_{\gamma\beta}
+ \nabla^{(0)}_\beta h_{\tau\gamma} - \nabla^{(0)}_\gamma h_{\tau\beta} \right)
\nabla^{(0)}_\rho \left( U \left(\phi^{(0)} \right)
R^{(0)\, \beta}_{\ \ \ \ \ \alpha\zeta\eta} \right) \nn
& - g^{(0)\, \beta\gamma} \left( \nabla^{(0)}_\tau h_{\gamma\alpha}
+ \nabla^{(0)}_\alpha h_{\tau\gamma} - \nabla^{(0)}_\gamma h_{\tau\alpha} \right)
\nabla^{(0)}_\rho
\left( U \left( \phi^{(0)} \right) R^{(0)\, \sigma}_{\ \ \ \ \ \beta\zeta\eta} \right) \nn
& - g^{(0)\, \beta\gamma} \left( \nabla^{(0)}_\tau h_{\gamma\zeta}
+ \nabla^{(0)}_\zeta h_{\tau\gamma} - \nabla^{(0)}_\gamma h_{\tau\zeta} \right)
\nabla^{(0)}_\rho
\left( U \left( \phi^{(0)} \right) R^{(0)\, \sigma}_{\ \ \ \ \ \alpha\beta\eta} \right) \nn
& \left. \left. - g^{(0)\, \beta\gamma} \left( \nabla^{(0)}_\tau h_{\gamma\eta}
+ \nabla^{(0)}_\eta h_{\tau\gamma} - \nabla^{(0)}_\gamma h_{\tau\eta} \right)
\nabla^{(0)}_\rho
\left( U \left( \phi^{(0)} \right) R^{(0)\, \sigma}_{\ \ \ \ \ \alpha\zeta\beta} \right) \right)
\right\}=0 \, .
\end{align}
In the Appendix, we  present the explicit form of the $(t,t)$, $(i,j)$,
and $(t,i)$ components
of the Einstein tensor of Eq.~(\ref{CS15}) in the FRW background (\ref{FRWmetric}).
We now choose the following gauge condition,
\begin{equation}
\label{CS17}
0 = \nabla^\mu h_{\mu\nu}\,  .
\end{equation}
In the FRW background (\ref{FRWmetric}), the gauge condition (\ref{CS17})
has the following forms,
\begin{align}
\label{gt}
0=& -3H h_{tt} -\frac{\partial h_{tt}}{\partial t}
 -\frac{H}{a^2}\left( h_{xx} +h_{yy} +h_{zz} \right)
+\frac{1}{a^2} \left( \frac{\partial h_{tx}}{\partial x}
+\frac{\partial h_{ty}}{\partial y} +\frac{\partial h_{tz}}{\partial z} \right)\, , \\
\label{gx}
0=& -3H h_{tx} -\frac{\partial h_{tx}}{\partial t}
+\frac{1}{a^2} \left( \frac{\partial h_{xx}}{\partial x}
+\frac{\partial h_{xy}}{\partial y} +\frac{\partial h_{xz}}{\partial z} \right)\, , \\
\label{gy}
0=& -3H h_{ty} -\frac{\partial h_{ty}}{\partial t}
+\frac{1}{a^2} \left( \frac{\partial h_{yy}}{\partial y}
+\frac{\partial h_{xy}}{\partial x} +\frac{\partial h_{yz}}{\partial z} \right)\, , \\
\label{gz}
0=& -3H h_{tz} -\frac{\partial h_{tz}}{\partial t}
+\frac{1}{a^2} \left( \frac{\partial h_{zz}}{\partial z}
+\frac{\partial h_{xz}}{\partial x} +\frac{\partial h_{yz}}{\partial y} \right)\, .
\end{align}
As we are interested in the gravitational wave, which corresponds to the massless
spin 2 mode, we assume that,
\begin{equation}
\label{CSg1}
h_{t\mu}=h_{\mu t}=0\, , \quad \sum_{i=1,2,3} h_{ii}=0 \, .
\end{equation}
Then Eqs.~(\ref{gt}), (\ref{gx}), (\ref{gy}), and (\ref{gz}) yield,
\begin{equation}
\label{CSg2}
0 = \sum_j \partial_j h_{ji}\, .
\end{equation}
In effect, Eq.~(\ref{tt}) in the Appendix indicates that
$G_{tt}=G_{tx}=G_{ty}=G_{tz}=0$, that is, $(t,t)$, $(t,x)$, $(t,y)$, and
$(t,z)$-components of (\ref{CS15}) are satisfied.
In addition, the other Einstein tensor components are,
\begin{align}
\label{xx2}
G_{xx}=& \dot H h_{xx} +H^2 h_{xx}  - \frac{1}{2} \frac{\partial^2 h_{xx} }{\partial t^2}
+H \frac{1}{2} \frac{\partial h_{xx} }{\partial t}
+\frac{1}{a^2} \left( \frac{\partial^2 h_{yz}}{\partial z \partial y}
 - \frac{1}{2} \frac{\partial^2 h_{yy}}{\partial z^2}
 - \frac{1}{2} \frac{\partial^2 h_{zz}}{\partial y^2} \right) \nn
& +\frac{8 \dot H \dot U}{a} \left( \frac{\partial h_{xz}}{\partial y}
-\frac{\partial h_{xy}}{\partial z}\right)
+H \left[ \frac{8 \ddot U}{a} \left( \frac{\partial h_{xz}}{\partial y}
 -\frac{\partial h_{xy}}{\partial z}\right)
+\frac{8 \dot U}{a} \left( \frac{\partial^2 h_{xz}}{\partial y \partial t}
 -\frac{\partial^2 h_{xy}}{\partial z \partial t}\right)  \right] \nn
& +\frac{4 \dot U}{a^3} \left( \frac{\partial^3 h_{xz}}{ \partial y^3}
 -\frac{\partial^3 h_{xy}}{ \partial z^3}
 -\frac{\partial^3 h_{xy}}{\partial z \partial y^2}
+\frac{\partial^3 h_{xz}}{\partial y \partial z^2}
 -\frac{\partial^3 h_{yz}}{\partial x \partial y^2}
+\frac{\partial^3 h_{yz}}{\partial x \partial z^2}
+\frac{\partial^3 (h_{yy}-h_{zz})}{\partial x \partial y \partial z}\right) \nn
& +\frac{4 \ddot U}{a} \left( \frac{\partial^2 h_{xy}}{\partial z \partial t}
 -\frac{\partial^2 h_{xz}}{\partial y \partial t}  \right)
+\frac{4 \dot U}{a} \left( -\frac{\partial^3 h_{xz}}{\partial y \partial t^2}
+\frac{\partial^3 h_{xy}}{\partial z \partial t^2} \right)\, , \\
\label{yy2}
G_{yy}=& \dot H h_{yy} +H^2 h_{yy}  - \frac{1}{2} \frac{\partial^2 h_{yy} }{\partial t^2}
+H \frac{1}{2} \frac{\partial h_{yy} }{\partial t}
+\frac{1}{a^2} \left( \frac{\partial^2 h_{zx}}{\partial x \partial z}
 - \frac{1}{2} \frac{\partial^2 h_{zz}}{\partial x^2}
 - \frac{1}{2} \frac{\partial^2 h_{xx}}{\partial z^2} \right) \nn
& +\frac{8 \dot H \dot U}{a} \left( \frac{\partial h_{yx}}{\partial z}
-\frac{\partial h_{yz}}{\partial x}\right)
+H \left[ \frac{8 \ddot U}{a} \left( \frac{\partial h_{yx}}{\partial z}
 -\frac{\partial h_{yz}}{\partial x}\right)
+\frac{8 \dot U}{a} \left( \frac{\partial^2 h_{yx}}{\partial z \partial t}
 -\frac{\partial^2 h_{yz}}{\partial x \partial t}\right)  \right] \nn
& +\frac{4 \dot U}{a^3} \left( \frac{\partial^3 h_{yx}}{ \partial z^3}
 -\frac{\partial^3 h_{yz}}{ \partial x^3}
 -\frac{\partial^3 h_{yz}}{\partial x \partial z^2}
+\frac{\partial^3 h_{yx}}{\partial z \partial x^2}
 -\frac{\partial^3 h_{zx}}{\partial y \partial z^2}
+\frac{\partial^3 h_{zx}}{\partial y \partial x^2}
+\frac{\partial^3 (h_{zz}-h_{xx})}{\partial y \partial z \partial x}\right) \nn
& +\frac{4 \ddot U}{a} \left( \frac{\partial^2 h_{yz}}{\partial x \partial t}
 -\frac{\partial^2 h_{yx}}{\partial z \partial t}  \right)
+\frac{4 \dot U}{a} \left( -\frac{\partial^3 h_{yx}}{\partial z \partial t^2}
+\frac{\partial^3 h_{yz}}{\partial x \partial t^2} \right)\, , \\
\label{zz2}
G_{zz}=& \dot H h_{zz} +H^2 h_{zz}  - \frac{1}{2} \frac{\partial^2 h_{zz} }{\partial t^2}
+H \frac{1}{2} \frac{\partial h_{zz} }{\partial t}
+\frac{1}{a^2} \left( \frac{\partial^2 h_{xy}}{\partial y \partial x}
 - \frac{1}{2} \frac{\partial^2 h_{xx}}{\partial y^2}
 - \frac{1}{2} \frac{\partial^2 h_{yy}}{\partial x^2} \right) \nn
& +\frac{8 \dot H \dot U}{a} \left( \frac{\partial h_{zy}}{\partial x}
-\frac{\partial h_{zx}}{\partial y}\right)
+H \left[ \frac{8 \ddot U}{a} \left( \frac{\partial h_{zy}}{\partial x}
 -\frac{\partial h_{zx}}{\partial y}\right)
+\frac{8 \dot U}{a} \left( \frac{\partial^2 h_{zy}}{\partial x \partial t}
 -\frac{\partial^2 h_{zx}}{\partial y \partial t}\right)  \right] \nn
& +\frac{4 \dot U}{a^3} \left( \frac{\partial^3 h_{zy}}{ \partial x^3}
 -\frac{\partial^3 h_{zx}}{ \partial y^3}
 -\frac{\partial^3 h_{zx}}{\partial y \partial x^2}
+\frac{\partial^3 h_{zy}}{\partial x \partial y^2}
 -\frac{\partial^3 h_{xy}}{\partial z \partial x^2}
+\frac{\partial^3 h_{xy}}{\partial z \partial y^2}
+\frac{\partial^3 (h_{xx}-h_{yy})}{\partial z \partial x \partial y}\right) \nn
& +\frac{4 \ddot U}{a} \left( \frac{\partial^2 h_{zx}}{\partial y \partial t}
 -\frac{\partial^2 h_{zy}}{\partial x \partial t}  \right)
+\frac{4 \dot U}{a} \left( -\frac{\partial^3 h_{zy}}{\partial x \partial t^2}
+\frac{\partial^3 h_{zx}}{\partial y \partial t^2} \right)\, , \\
\label{xy2}
G_{xy}=& \dot H \left[ h_{xy}
+\frac{4 \dot U}{a} \left( \frac{\partial (h_{xx}-h_{yy})}{\partial z}
+\frac{\partial h_{yz}}{\partial y} -\frac{\partial h_{xz}}{\partial x} \right) \right]
+H^2 h_{xy} \nn
& +H \left[ \frac{1}{2} \frac{\partial h_{xy}}{\partial t}
+\frac{4 \ddot U}{a}\left( \frac{\partial (h_{xx}-h_{yy})}{\partial z}
+\frac{\partial h_{yz}}{\partial y} -\frac{\partial h_{xz}}{\partial x} \right)
\right. \nn & \left.
+\frac{4 \dot U}{a}\left( \frac{\partial^2 (h_{xx}-h_{yy})}{\partial z \partial t}
+\frac{\partial^2 h_{yz}}{\partial y \partial t}
 -\frac{\partial^2 h_{xz}}{\partial x \partial t} \right)\right]
 - \frac{1}{2} \frac{\partial^2 h_{xy}}{\partial t^2} \nn
& +\frac{2 \ddot U}{a}\left( \frac{\partial^2 (h_{yy}-h_{xx})}{\partial z \partial t}
+\frac{\partial^2 h_{xz}}{\partial x \partial t}
 -\frac{\partial^2 h_{yz}}{\partial y \partial t} \right)
+\frac{2 \dot U}{a}\left( \frac{\partial^3 (h_{yy} - h_{xx})}{\partial z \partial t^2}
+\frac{\partial^3 h_{xz}}{\partial x \partial t^2}
 -\frac{\partial^3 h_{yz}}{\partial y \partial t^2} \right) \nn &
+\frac{1}{2 a^2} \left( \frac{\partial^2 h_{xy}}{\partial z^2}
+\frac{\partial^2 h_{zz}}{\partial y \partial x}
 -\frac{\partial^2 h_{xz}}{\partial z \partial y}
 -\frac{\partial^2 h_{yz}}{\partial z \partial x} \right)
+\frac{2 \dot U}{a^3}\left( \frac{\partial^3 (h_{xx}-h_{yy})}{\partial z^3}
+\frac{\partial^3 (h_{zz}-h_{yy})}{\partial z \partial x^2} \right. \nn & \left.
+\frac{\partial^3 (h_{xx}-h_{zz})}{\partial z \partial y^2}
+2\frac{\partial^3 h_{yz}}{\partial y \partial x^2}
 -2\frac{\partial^3 h_{xz}}{\partial x \partial y^2}
+2\frac{\partial^3 h_{yz}}{\partial y \partial z^2}
 -2\frac{\partial^3 h_{xz}}{\partial x \partial z^2}  \right)\, , \\
\label{xz2}
G_{xz}=& \dot H \left[ h_{xz} -\frac{4 \dot U}{a}
\left( \frac{\partial (h_{xx}-h_{zz})}{\partial y}
+\frac{\partial h_{yz}}{\partial z} -\frac{\partial h_{xy}}{\partial x} \right) \right]
+H^2 h_{xz} \nn
& +H \left[ \frac{1}{2} \frac{\partial h_{xz}}{\partial t}
 -\frac{4 \ddot U}{a}\left( \frac{\partial (h_{xx}-h_{zz})}{\partial y}
+\frac{\partial h_{yz}}{\partial z} -\frac{\partial h_{xy}}{\partial x} \right)
\right. \nn & \left.
 -\frac{4 \dot U}{a}\left( \frac{\partial^2 (h_{xx}-h_{zz})}{\partial y \partial t}
+\frac{\partial^2 h_{yz}}{\partial z \partial t}
 -\frac{\partial^2 h_{xy}}{\partial x \partial t} \right)\right]
 - \frac{1}{2} \frac{\partial^2 h_{xz}}{\partial t^2} \nn
& -\frac{2 \ddot U}{a}\left( \frac{\partial^2 (h_{zz}-h_{xx})}{\partial y \partial t}
+\frac{\partial^2 h_{xy}}{\partial x \partial t}
 -\frac{\partial^2 h_{yz}}{\partial z \partial t} \right)
 -\frac{2 \dot U}{a}\left( \frac{\partial^3 (h_{zz} -h_{xx})}{\partial y \partial t^2}
+\frac{\partial^3 h_{xy}}{\partial x \partial t^2}
 -\frac{\partial^3 h_{yz}}{\partial z \partial t^2} \right) \nn
& +\frac{1}{2 a^2} \left( \frac{\partial^2 h_{xz}}{\partial y^2}
+\frac{\partial^2 h_{yy}}{\partial z \partial x}
 -\frac{\partial^2 h_{xy}}{\partial z \partial y}
 -\frac{\partial^2 h_{yz}}{\partial y \partial x} \right)
 -\frac{2 \dot U}{a^3}\left( \frac{\partial^3 (h_{xx}-h_{zz})}{\partial y^3}
+\frac{\partial^3 (h_{yy}-h_{zz})}{\partial y \partial x^2}
\right. \nn & \left.
+\frac{\partial^3 (h_{xx}-h_{yy})}{\partial y \partial z^2}
+2\frac{\partial^3 h_{yz}}{\partial z \partial x^2}
 -2\frac{\partial^3 h_{xy}}{\partial x \partial z^2}
+2\frac{\partial^3 h_{yz}}{\partial z \partial y^2}
 -2\frac{\partial^3 h_{xy}}{\partial x \partial y^2}  \right)\, , \\
\label{yz2}
G_{yz}=& \dot H \left[ h_{yz} +\frac{4 \dot U}{a}
\left( \frac{\partial (h_{yy}-h_{zz})}{\partial x}
+\frac{\partial h_{xz}}{\partial z} -\frac{\partial h_{xy}}{\partial y} \right) \right]
+H^2 h_{yz} \nn
& +H \left[ \frac{1}{2} \frac{\partial h_{yz}}{\partial t}
+\frac{4 \ddot U}{a}\left( \frac{\partial (h_{yy}-h_{zz})}{\partial x}
+\frac{\partial h_{xz}}{\partial z} -\frac{\partial h_{xy}}{\partial y} \right)
\right. \nn
& \left. +\frac{4 \dot U}{a}\left( \frac{\partial^2 (h_{yy}-h_{zz})}{\partial x \partial t}
+\frac{\partial^2 h_{xz}}{\partial z \partial t}
 -\frac{\partial^2 h_{xy}}{\partial y \partial t} \right)\right]
 - \frac{1}{2} \frac{\partial^2 h_{yz}}{\partial t^2} \nn
& +\frac{2 \ddot U}{a}\left( \frac{\partial^2 (h_{zz}-h_{yy})}{\partial x \partial t}
+\frac{\partial^2 h_{xy}}{\partial y \partial t}
 -\frac{\partial^2 h_{xz}}{\partial z \partial t} \right)
+\frac{2 \dot U}{a}\left( \frac{\partial^3 (h_{zz}-h_{yy})}{\partial x \partial t^2}
+\frac{\partial^3 h_{xy}}{\partial y \partial t^2}
 -\frac{\partial^3 h_{xz}}{\partial z \partial t^2} \right) \nn
& +\frac{1}{2 a^2} \left( \frac{\partial^2 h_{yz}}{\partial x^2}
+\frac{\partial^2 h_{xx}}{\partial z \partial y}
 -\frac{\partial^2 h_{xy}}{\partial z \partial x}
 -\frac{\partial^2 h_{xz}}{\partial y \partial x} \right)
+\frac{2 \dot U}{a^3}\left( \frac{\partial^3 (h_{yy}-h_{zz})}{\partial x^3}
+\frac{\partial^3 (h_{xx}-h_{zz})}{\partial x \partial y^2}
\right. \nn & \left.
+\frac{\partial^3 (h_{yy}-h_{xx})}{\partial x \partial z^2}
+2\frac{\partial^3 h_{xz}}{\partial z \partial y^2}
 -2\frac{\partial^3 h_{xy}}{\partial y \partial z^2}
+2\frac{\partial^3 h_{xz}}{\partial z \partial x^2}
 -2\frac{\partial^3 h_{xy}}{\partial y \partial x^2}  \right)\, .
\end{align}
In order to reveal the propagation properties at present time, let us assume that
the variations of the Hubble rate $H$ and of the axion Chern-Simons coupling function
$\dot U$ during the propagation of the
gravitational wave are negligible and they are constant, that is,
\begin{equation}
\label{CGRa1}
H^2=H_0^2\, , \quad \dot H = H_1\, , \quad \dot U = U_0\, , \quad \ddot U = 0 \, ,
\end{equation}
with constants $H_0$, $H_1$, and $U_0$. We also assume that the scale factor $a$ is slowly
varying at present time, so we shall assume that it is approximately equal to
unity, $a\simeq 1$. We also consider the gravitational wave propagating along
the $z$-direction, that is,
$h_{ij} \propto \e^{-i\left(\omega t - k z \right)}$, with a constant angular frequency
$\omega$ and a constant wavenumber $k$. Then the conditions in (\ref{CSg1}) and
(\ref{CSg2}) indicate that,
\begin{equation}
\label{CGRa2}
h_{iz}=0\, , \quad h_{xx}=- h_{yy} = h_+ \e^{-i\left(\omega t - k z \right)} \, , \quad
h_{xy}=h_{yx} = h_\times \e^{-i\left(\omega t - k z \right)} \, ,
\end{equation}
with complex constants $h_+$ and $h_\times$, which express the polarizations
of the gravitational wave. Only the real parts in (\ref{CGRa2}) are physically meaningful. Then the non-zero components of the Einstein tensor are,
\begin{align}
\label{xxyy3}
G_{xx}=& - G_{yy} \nn
=& \left( H_1 +H_0^2 - \frac{i\omega H_0}{2} + \frac{\omega^2}{2}
 - \frac{k^2}{2} \right) h_+
 -4 ik \left( 2 H_1 - 2 i \omega H_0 - k^2 + \omega^2 \right) U_0 h_\times \, , \\
\label{xy3}
G_{xy}=& \left( H_1 + H_0^2 - \frac{i\omega H_0}{2} + \frac{\omega^2}{2} - \frac{k^2}{2}
\right) h_\times
+ 4 ik \left( 2 H_1 - 2 i \omega H_0 -  k^2 + \omega^2 \right)U_0 h_+ \, , \\
\label{zzxzyz}
G_{zz}=& G_{xz} = G_{yz} = 0 \, .
\end{align}
An interesting situation arises when $U_0\neq 0$ and $H\neq 0$
$\left(H_0\neq 0\ \mbox{or}\,\, H_1\neq 0\right)$, in which case there is always
a mixing of $+$-mode
corresponding to $h_+$ and $\times$-mode to $h_\times$.
This scenario is particularly interesting since there is a non-trivial polarization mode
for the gravitational wave, and also the dispersion relation for the two polarization
modes corresponding to the Right-handed mode $h_\times = i h_+$ and the Left-handed mode
$h_\times = - i h_+$ are different from each other.
We should note, however, even if $U_0\neq 0$, in a flat background
where $H=0$ $\left(H_0=H_1=0\right)$,
as long as the dispersion relation $\omega^2=k^2$ is satisfied, the $+$-mode and
$\times$-mode becomes 
independent from each other.

We now explain the above mentioned things in more details. From the condition that the equations $G_{xx}=G_{yy}=G_{xy}=0$ have non-trivial solutions, we find the dispersion relation,
\begin{equation}
\label{CGd1}
\left( H_1 +H_0^2 - \frac{i\omega H_0}{2} + \frac{\omega^2}{2}
 - \frac{k^2}{2} \right)^2
= 16 k^2 \left( 2 H_1 - 2 i \omega H_0 + \omega^2 - k^2 \right)^2 U_0^2 \, ,
\end{equation}
that is,
\begin{equation}
\label{CGd2}
H_1 +H_0^2 - \frac{i\omega H_0}{2} + \frac{\omega^2}{2}
 - \frac{k^2}{2}
= \pm 4 k \left( 2 H_1 - 2 i \omega H_0 + \omega^2 - k^2 \right) U_0 \, ,
\end{equation}
which indicates that $h_\times = \pm i h_+$ and if $k>0$, $+$ sign $h_\times = i h_+$ corresponds
to the Right-handed polarization and $-$ sign
$h_\times = i h_+$ to the Left-handed polarization (see \cite{Seto:2008sr}, for example).
Therefore the Right-handed polarization mode has a dispersion relation different from that
of the Left-handed polarization.
The dispersion relation (\ref{CGd2}) can be rewritten in the following form,
\begin{equation}
\label{CGd3}
0= D_{^R_L} \left( \omega, k \right)
\equiv \omega^2 - k^2 + \frac{2 H_1 + 2 H_0^2 - i\omega H_0
\mp 16kU_0 \left( H_1 - 2i \omega H_0 \right)}{1 \mp 8k U_0} \, .
\end{equation}
If we assume $\left| kU_0 \right| \ll 1$, Eq.~(\ref{CGd3}) has the following form,
\begin{equation}
\label{CGd3}
0= D_{^R_L} \left( \omega, k \right)
\sim \omega^2 - k^2 + 2 H_1 + 2 H_0^2 - i\omega H_0
\pm 16 kU_0 H_0^2 \pm 24 i \omega k U_0 H_0 \, .
\end{equation}
Then although the term $ - i\omega H_0$ gives the dissipation coming
from the expansion of the universe, the last term $\pm 24 i \omega k U_0 H_0$ works
against for the dissipation in the Right-handed mode (now we assume $k>0$)
but works to increase the dissipation for the Left-handed mode.
When we consider the high frequency mode $\left|kU_0\right| \gg 1$
and $\omega H_0 \gg H_1$, Eq.~(\ref{CGd3}) has the following form,
\begin{equation}
\label{CGd3}
0= D_{^R_L} \left( \omega, k \right)
\sim \omega^2 - k^2 - 4i \omega H_0 \, .
\end{equation}
We may compare the above expression with the expression in case $U_0=0$
and for the high frequency mode,
\begin{equation}
\label{CGd4}
0= D_{^R_L} \left( \omega, k \right)
\sim \omega^2 - k^2 - i \omega H_0 \, .
\end{equation}
Although $U_0$ does not appear in the expression of Eq.~(\ref{CGd3}), the dissipation is
four times stronger than that in case of $U_0=0$ in (\ref{CGd4}).

In general relativity, the gravitational wave has two modes corresponding to
the helicity, that is,
the Right-handed mode and the Left handed mode.
The two modes have the identical dispersion relation, which enables to consider
$+$-mode and $\times$-mode, instead of the Right- and Left-handed modes.
In the model which we are now consider, we cannot consider the $+$-mode and
the $\times$-mode as independent modes because the Right- and Left-handed modes
satisfy different dispersion relations, respectively.
This is because the model breaks parity and therefore the model is chiral, which is an important result of this work. This polarization asymmetry of the two propagating modes of the gravitational wave could be detected in the future in the LIGO or the forthcoming LISA collaborations. In addition, the polarization of the gravitational wave in the early Universe also affects the polarization of CMB, and specifically the E-mode and B-modes, see for example \cite{Bielefeld:2014nza}.

\section{Gravitational Waves in $F(R)$ gravity}

For comparison reasons, in this section we shall study the gravitational waves in
$F(R)$ gravity, with the action being,
\begin{equation}
\label{FRaction}
S=\int d^4x\sqrt{-g}\left[
\frac{F(R)}{2\kappa^2} + \mathcal{L}_\mathrm{matter} \left( g_{\mu\nu}, \Phi_i \right)
\right]\, ,
\end{equation}
where $\mathcal{L}_\mathrm{matter} $ is the Lagrangian density
of the matter fluids present, and the
$\Phi_i$'s express the various different matter fields.
We can rewrite the action~(\ref{FRaction}) by introducing the
auxiliary scalar fields $A$ and $B$ in the following way,
\begin{equation}
\label{FhGV2}
S_{AB} = \int d^4 x \sqrt{-g} \left[ \frac{1}{2\kappa^2} \left\{ B \left( R - A \right)
+ F(A) \right\}
+ \mathcal{L}_\mathrm{matter} \left( g_{\mu\nu}, \Phi_i \right) \right] \, .
\end{equation}
Upon variation of the action with respect to the auxiliary scalar
$A$, we obtain,
\begin{equation}
\label{FhGV3}
B = F'(A) \, .
\end{equation}
By redefining the scalar field $B$ by introducing a new scalar field $\sigma$ with
$B=\e^{\sigma}$, we assume that
Eq.~(\ref{FhGV3}) can be solved with respect to $A$ as $A=A(\sigma)$.
In effect, the action~(\ref{FhGV2}) can be rewritten as
follows,
\begin{align}
\label{FhGV5}
S_{\sigma\phi} = \int d^4 x \sqrt{-g} \left[ \frac{1}{2\kappa^2} \left\{ \e^\sigma
\left( R - A\left(\sigma\right) \right)
+ F\left(A \left(\sigma\right) \right) \right\}
+ \mathcal{L}_\mathrm{matter} \left( g_{\mu\nu}, \Phi_i \right) \right] \, .
\end{align}
By performing a scale transformation of the metric tensor,
\begin{equation}
\label{FhGV6}
g_{\mu\nu} = \e^{-\sigma} {\tilde g}_{\mu\nu} \, ,
\end{equation}
the action~(\ref{FhGV5}) can be transformed in the Einstein frame,
and it is equal to,
\begin{align}
\label{FhGV7} S_\mathrm{E} =& \int d^4 x \sqrt{-\tilde g} \left[
\frac{1}{2\kappa^2} \left\{ \tilde R - \frac{3}{2} \partial_\mu
\sigma \partial^\mu \sigma \right\}  - U \left(\sigma\right) +
\e^{-2\sigma} \mathcal{L}_\mathrm{matter} \left( \e^{-\sigma}
{\tilde g}_{\mu\nu}, \Phi_i \right) \right] \, , \nn U
\left(\sigma\right) \equiv & \e^{-\sigma} A\left(\sigma\right)
 - \e^{-2\sigma} F\left(A \left(\sigma\right) \right) \, .
\end{align}
Let us now consider the gravitational wave, based on the action~(\ref{FhGV7})
in the Einstein frame by considering the
perturbation of the background metric, ${\tilde g}_{\mu\nu} = {\tilde
g^{(0)}}_{\mu\nu}$ as ${\tilde g}_{\mu\nu} = {\tilde
g^{(0)}}_{\mu\nu} + {\tilde h}_{\mu\nu}$ in the Einstein equation, so we have,
\begin{equation}
\label{FhGV16}
{\tilde R}_{\mu\nu} - \frac{1}{2} {\tilde g}_{\mu\nu} {\tilde R}
= 3 \partial_\mu \sigma \partial_\nu \sigma
+ {\tilde g}_{\mu\nu} \left( - \frac{3}{2} \partial_\rho \sigma \partial^\rho \sigma
 - U \left( \sigma \right) \right)
+ \kappa^2 {\tilde T}_{\mathrm{matter}\, \mu\nu} \, .
\end{equation}
Here the matter energy momentum tensor ${\tilde
T}_{\mathrm{matter}}^{\mu\nu}$ in the Einstein frame is,
\begin{equation}
\label{FhGV17O}
{\tilde T}_{\mathrm{matter}}^{\mu\nu} \equiv \frac{2}{\sqrt{- \tilde g}}
\frac{\partial \left( \sqrt{- \tilde g} \e^{-2\sigma} \mathcal{L}_\mathrm{matter}
\left( \e^{-\sigma} {\tilde g}_{\mu\nu}, \Phi_i \right) \right)}
{\partial {\tilde g}_{\mu\nu}} \, .
\end{equation}
If the matter fluids are minimally coupled with gravity, that is, if
the matter Lagrangian is of the form $ \mathcal{L}_\mathrm{matter} \left(
\e^{-\sigma} {\tilde g}_{\mu\nu}, \Phi_i \right) =
\mathcal{L}_\mathrm{matter} \left( g_{\mu\nu}, \Phi_i \right)$,
so it does not include any derivative of the metric $g_{\mu\nu}$, the
matter energy momentum tensor ${\tilde
T}_{\mathrm{matter}}^{\mu\nu}$ in the Einstein frame is related
with the matter energy momentum tensor
$T_{\mathrm{matter}}^{\mu\nu}$ in the original Jordan frame as
${\tilde T}_{\mathrm{matter}}^{\mu\nu} = \e^{-3\sigma}
T_{\mathrm{matter}}^{\mu\nu}$, that is, ${\tilde
T}_{\mathrm{matter}\, \mu\nu} = \e^{-\sigma} T_{\mathrm{matter}\,
\mu\nu}$. When we consider the gravitational wave, we often use
the transverse and traceless gauge conditions. Since we are
considering the scale transformation~(\ref{FhGV6}), if
$h_{\mu\nu}$, which is defined as the fluctuation from the
background metric $g_{\mu\nu} = g^{(0)}_{\mu\nu}$ as $g_{\mu\nu} =
g^{(0)}_{\mu\nu} + h_{\mu\nu}$ in the original frame in
the action~(\ref{FRaction}), it satisfies the transverse and traceless gauge
conditions,
\begin{equation}
\label{FhGV17} \nabla^\mu h_{\mu\nu} = g^{(0)\, \mu\nu} h_{\mu\nu}
= 0 \, .
\end{equation}
However, the scale transformed fluctuation ${\tilde h}_{\mu\nu} =
\e^\sigma h_{\mu\nu}$ does not always satisfy the first
condition in Eq.~(\ref{FhGV17}), although the second condition is
trivially satisfied, ${\tilde g}^{(0)\, \mu\nu} {\tilde h}_{\mu\nu}
= \e^{-\sigma} g^{(0)\, \mu\nu} \e^\sigma h_{\mu\nu}
= g^{(0)\, \mu\nu} h_{\mu\nu} = 0$. For the first condition in
Eq.~(\ref{FhGV17}), under the scale transformation, we find,
\begin{align}
\label{FhGV18}
& \tilde{\nabla}^\mu \tilde{h}_{\mu}^{\ \nu}
= \e^{-\sigma} \nabla^\mu h_{\mu\nu}
+ 4 \e^{-\sigma} g^{(0)\, \mu\tau} g^{(0)\, \nu\rho} \sigma_{,\tau} h_{\mu\rho}
 - \e^{-\sigma} g^{(0)\, \nu\rho}\sigma_{,\rho} g^{(0)\, \mu\tau} h_{\mu\tau}
= 4 \e^{-\sigma} g^{(0)\, \mu\tau} g^{(0)\, \nu\rho} \sigma_{,\tau} h_{\mu\rho} \, .
\end{align}
Then if we assume that we have a homogeneous and isometric background
metric, $\sigma$ depends solely on the cosmological
time $t$, and also $g^{(0)}_{ti}=0$. In effect, if we consider the
perturbation with $h_{t\mu}=0$ since we are considering the
massless spin 2 mode, we find,
\begin{equation}
\label{FhGV19} \tilde{\nabla}^\mu \tilde{h}_{\mu}^{\ \nu} =
{\tilde g}^{(0)\, \mu\nu} {\tilde h}_{\mu\nu} = 0\, .
\end{equation}
Therefore, the gauge conditions in (\ref{FhGV17}) for the graviton are not changed
by the scale transformation under consideration,
\begin{equation}
\label{FhGV21}
{\tilde\nabla}^\mu {\tilde h}_{\mu\nu} = {\tilde g}^{(0)\, \mu\nu} {\tilde h}_{\mu\nu} = 0 \, .
\end{equation}
Then under the condition~(\ref{FhGV21}), the equation for the
gravitational wave can be written as follows,
\begin{align}
\label{FhGV22}
0 =& \frac{1}{2\kappa^2}\left(- \frac{1}{2}\left( - {\tilde \Box}^{(0)} {\tilde h}_{\mu\nu}
 - 2{\tilde R}^{(0)\, \lambda\ \rho}_{\ \ \ \ \ \nu\ \mu} {\tilde h}_{\lambda\rho}
+ {\tilde R}^{(0)\, \rho}_{\ \ \ \ \ \mu}{\tilde h}_{\rho\nu}
+ {\tilde R}^{(0)\, \rho}_{\ \ \ \ \ \nu}{\tilde h}_{\rho\mu} \right)
+ \frac{1}{2} R^{(0)} h_{\mu\nu}
 - \frac{1}{2}{\tilde g}^{(0)}_{\mu\nu} {\tilde h}_{\rho\sigma}
{\tilde R}^{(0)\, \rho\sigma} \right) \nn
& + {\tilde h}_{\mu\nu} \left( - \frac{3}{2} \partial_\rho \sigma \partial^\rho \sigma
 - U \left( \sigma \right) \right)
+ \frac{3}{2} {\tilde g}^{(0)}_{\mu\nu} \partial^\rho \sigma \partial^\tau \sigma
{\tilde h}_{\rho\tau}
+ \frac{1}{2} \frac{\partial {\tilde T}_{\mathrm{matter}\, \mu\nu}}
{\partial {\tilde g}_{\rho\tau}}
{\tilde h}_{\rho\tau} \, .
\end{align}
We are now interested in the massless spin-two mode, which
satisfies,
\begin{equation}
\label{FhGV23}
{\tilde h}_{it}={\tilde h}_{ti}=h_{it}=h_{ti}=0\, , \quad
\sum_{i=1,2,3} {\tilde h}_{ii}=\sum_{i=1,2,3} h_{ii}=0 \, , \quad i=1,2,3\, ,
\quad {\tilde h}_{tt}=h_{tt}=0 \, .
\end{equation}
In the spatially flat FRW universe in the Einstein frame,
\begin{equation}
\label{JGRG14E}
d{\tilde s}^2 \equiv \e^{\sigma} ds^2 = - d{\tilde t}^2
+ {\tilde a} \left( \tilde t \right)^2 \sum_{i=1,2,3} \left(dx^i\right)^2\, ,
\end{equation}
where $d{\tilde t}\equiv \e^{\frac{\sigma}{2}} dt$ and $\tilde a
\left( \tilde t \right) \equiv \e^{\frac{\sigma}{2}} a(t)$, due to
the isometry in the spacial part, we may assume,
\begin{equation}
\label{FhGV24}
\frac{\partial {\tilde T}_{\mathrm{matter}\, tt}}{\partial{\tilde g}_{ij}} \propto \delta^{ij} \, , \quad
\frac{\partial {\tilde T}_{\mathrm{matter}\, tk}}{\partial {\tilde g}_{ij}}
= \frac{\partial {\tilde T}_{\mathrm{matter}\, kt}}{\partial {\tilde g}_{ij}} = 0 \, .
\end{equation}
We may further assume that the matter energy-momentum tensor~(\ref{FhGV17O})
in the Einstein frame has the following perfect fluid form,
\begin{equation}
\label{FhGV25}
{\tilde T}_{\mathrm{matter}\, \mu\nu} =\tilde\rho {\tilde U}_\mu {\tilde U}_\nu
+ \tilde p {\tilde\gamma}_{\mu\nu} \, .
\end{equation}
Here $\left( \tilde U^\mu \right)$ is the four velocity of the matter
fluid and we now assume $\tilde U^0=1$ and $\tilde U^i=0$. In
Eq.~(\ref{FhGV25}), ${\tilde\gamma}_{\mu\nu}$ is the projection
tensor to the spatial directions perpendicular to $\tilde U^\mu$,
\begin{equation}
\label{FhGV26}
{\tilde\gamma}_{\mu\nu}={\tilde g}_{\mu\nu} + {\tilde U}_\mu {\tilde U}_\nu\, .
\end{equation}
We now also assume that the matter fluid minimally couples with
the metric ${\tilde g}_{\mu\nu}$, that is, the coupling between
the matter fluids and the metric does not include the derivative
of the metric. Then, under the perturbation ${\tilde g}_{\mu\nu} =
{\tilde g}^{(0)}_{\mu\nu} + {\tilde h}_{\mu\nu}$, we find,
\begin{equation}
\label{FhGV27}
\delta \tilde \rho = {\tilde \rho}^{\mu\nu} {\tilde h}_{\mu\nu}\, , \quad
\delta \tilde p = {\tilde p}^{\mu\nu} {\tilde h}_{\mu\nu}\, .
\end{equation}
On the other hand, the variation of ${\tilde U}_\mu$ is given by
using the condition ${\tilde U}^\mu {\tilde U}_\mu = -1$, that is,
\begin{equation}
\label{FhGV29}
0 = 2 \left( \delta {\tilde U}^\mu \right)
+ {\tilde U}^\mu {\tilde U}^\nu {\tilde h}_{\mu\nu}
= {\tilde U}^\mu \left( 2 {\tilde g}^{(0)}_{\mu\nu} \delta {\tilde U}^\nu
+ {\tilde h}_{\mu\nu} {\tilde U}^\nu \right)\, ,
\end{equation}
and hence we have,
\begin{equation}
\label{FhGV31} \delta {\tilde U}^\mu = - \frac{1}{2} {\tilde
g}^{(0)\, \mu\rho} \left( {\tilde h}_{\rho\nu} {\tilde U}^\nu +
l_\rho \right)\, .
\end{equation}
Here $l_\mu$ is an arbitrary vector which satisfies the
condition ${\tilde U}^\mu l_\mu = 0$,
but we choose $l_\mu=0$ by assuming the isometry.
Due to the isometry, we may assume $\rho^{ij}$ and $p^{ij}$ are
proportional to $\delta^{ij}$, Then the $(t,t)$ and $(t,i)$
components of equation for the gravitational wave (\ref{FhGV22}) are
trivially satisfied and the $(i,j)$ component is given by,
\begin{equation}
\label{FhGV32}
0 = \frac{1}{2\kappa^2}\left( \frac{1}{2}
\left( - \partial_{\tilde t}^2 {\tilde h}_{ij} + {\tilde a}^{-2} \triangle {\tilde h}_{ij} \right)
+ \left( 3 \frac{d \tilde H}{d\tilde t} + 4 {\tilde H}^2 \right) {\tilde h}_{ij}
+ {\tilde h}_{ij} \left( \frac{3}{2} \left( \frac{d \sigma}{d\tilde t} \right)^2
 - U \left( \sigma \right) \right) \right) \, .
\end{equation}
We may rewrite Eq.~(\ref{FhGV32}) in the original Jordan frame, and since,
${\tilde h}_{\mu\nu} = \e^\sigma h_{\mu\nu}$ and
$\tilde a = \e^{\frac{\sigma}{2}} a$, we find the following,
\begin{align}
\label{FhGV33}
\frac{\partial}{\partial \tilde t} =& \e^{-\frac{\sigma}{2}} \frac{\partial}{\partial t} \, , \quad
\frac{\partial^2}{\partial \tilde t^2} = \e^{-\sigma}
\left( \frac{\partial^2}{\partial t^2 } - \frac{1}{2}\frac{d\sigma}{dt} \frac{\partial}{\partial t}
\right) \, , \nn
\tilde H =& \frac{1}{\tilde a} \frac{d \tilde a}{d\tilde t}
= \e^{-\frac{\sigma}{2}} \left( \frac{1}{2} \frac{d \sigma }{d t} + H \right) \, , \quad
\frac{d \tilde H}{d\tilde t} = \e^{-\tilde \sigma} \left( \frac{1}{2} \frac{d^2 \sigma}{dt^2}
+ \frac{1}{4} \left( \frac{d \sigma}{dt} \right)^2 + \frac{d H}{dt} \right) \, , \nn
\frac{\partial^2 {\tilde h}_{ij}}{\partial \tilde t^2} =&
\frac{\partial^2 h_{ij} }{\partial t^2} + \frac{d^2 \sigma}{dt^2} h_{ij}
+ \left( \frac{d\sigma}{dt} \right)^2 h_{ij}
+ 2 \frac{d \sigma}{dt} \frac{\partial h_{ij}}{\partial t}
 - \frac{1}{2} \left( \frac{d \sigma}{dt} \right)^2 \frac{\partial h_{ij}}{\partial t}
 - \frac{1}{2}\frac{d\sigma}{dt} \frac{\partial h_{ij}}{\partial t} \nn
=& \frac{\partial^2 h_{ij} }{\partial t^2} + \frac{d^2 \sigma}{dt^2} h_{ij}
+ \frac{1}{2} \left( \frac{d\sigma}{dt} \right)^2 h_{ij}
+ \frac{3}{2} \frac{d \sigma}{dt} \frac{\partial h_{ij}}{\partial t} \, .
\end{align}
Then, due to the fact that $\tilde p = \e^{-2\sigma} p$ from
 Eq.~(\ref{FhGV25}) (${\tilde T}_{\mathrm{matter}\, \mu\nu} =
\e^{-\sigma} T_{\mathrm{matter}\, \mu\nu}$ and
${\tilde\gamma}_{\mu\nu} = \e^\sigma \gamma_{\mu\nu}$),
Eq.~(\ref{FhGV32}) can be rewritten as follows,
\begin{align}
\label{FhGV34}
0 =& \frac{1}{2\kappa^2}\left( \frac{1}{2}
\left( - \partial_t^2 h_{ij} + \frac{3}{2} \dot\sigma \partial_t h_{ij}
+ \left( \ddot\sigma + {\dot\sigma}^2 \right) h_{ij} + a^{-2} \triangle h_{ij} \right) \right. \nn
& \left. + \left( 3 \dot H + 4 H^2 + \frac{3}{2} \ddot \sigma + \frac{13}{4} {\dot \sigma}^2
+ 4 \dot\sigma H - \e^{\sigma} U \left( \sigma \right) \right) h_{ij} \right)
+ \frac{1}{2}\e^{-\sigma} p h_{ij} \, .
\end{align}
We should note that $\e^\sigma$ is given in Eq.~(\ref{FhGV3}) and
$U \left( \sigma \right)$ is given in Eq.~(\ref{FhGV7}).
Then in terms of the Jordan frame, we find,
\begin{equation}
\label{FhGV35}
\sigma = \ln \left( F'(R) \right) \, , \quad
U \left( \sigma \right) = \e^{-\sigma} R - \e^{-2\sigma} F\left( R \right)
= \frac{R F'(R) - F\left( R \right) }{F'(R)^2} \, .
\end{equation}
From Eq.~(\ref{FhGV34}), it is clear that if $\dot\sigma > 0$,
the gravitational wave is enhanced and if $\dot\sigma <0$, dissipation of the gravity
wave occurs. The enhancement or the dissipation of the
gravity wave occurs due to the term
$\sim \frac{3}{2} \dot\sigma \partial_t h_{ij}$
in Eq.~(\ref{FhGV34}), which includes the first
derivative of $h_{ij}$. The enhancement or the dissipation of the gravity wave occurs as an
effect originating from the scale transformation,
${\tilde h}_{\mu\nu} = \e^\sigma h_{\mu\nu}$.

Comparing the propagation of the gravitational wave in $F(R)$ gravity
with the gravitational wave in the Einstein Chern-Simons gravity,
it is apparent that
in the $F(R)$ gravity case, there is no difference in the propagation between
the right-handed mode and the left-handed mode in helicity, an effect which occurs
due to the fact that there is no violation of parity
in $F(R)$ gravity. However it is expected that the presence of the Chern-Simons term
in vacuum $F(R)$ gravity may induce non-trivial phenomena.
We should also note that there appears a dissipation or enhancement term coming
from the Chern-Simons term in the Chern-Simons Einstein gravity, similar term also
appear in the $F(R)$ gravity.

\section{Conclusions}

In this work we studied the propagation of gravitational waves in the Chern-Simons
axion Einstein gravity context.
Our aim was to examine whether it is possible to reveal at a quantitative level,
any significant difference between the two polarization modes.
We performed the study assuming a flat FRW background, and by studying the tensor
perturbations of the metric, we demonstrated that the $+$-mode and $\times$-mode
cannot be independent from each other but they appear only as the combination of the
Right-handed mode or the Left-handed mode and
the dispersion relations of the Right-handed mode and the Left-handed mode
are different, even for the simplest form of the Chern-Simons
scalar axionic coupling. This is because the Chern-Simons scalar
term breaks parity and therefore the model is chiral. We may
expect that the difference of the polarization could be detected
by LIGO and forthcoming LISA collaboration. Also, the polarization
of the primordial gravity waves influences the polarization of
CMB, E-mode and B-mode in the early Universe
\cite{Bielefeld:2014nza}. It is conceivable that more complex
axion scalar couplings to the Chern-Simons term may further
perplex the dispersion relations for the two polarization modes.
For the low-frequency mode, the Chern-Simons scalar coupling works
against for the dissipation in the Right-handed mode works to
increase the dissipation for the Left- handed mode. On the other
hand, for the high frequency mode, the Chern-Simons scalar
coupling makes the dissipation of the gravitational wave four
times stronger than that in case of the standard Einstein gravity.
$U_0=0$ in (\ref{CGd4}). In addition to the above findings, we
demonstrated that there exists a non-trivial mixing between the
two different polarization modes, which strongly suggests
differences between the standard scalar axion Einstein gravity and
the Chern-Simons axion Einstein gravity. Just for comparison, we
also investigated the propagation of the gravitational wave in the
$F(R)$ gravity model and found that there is no difference between
the Right-handed mode and the Left-handed mode, which is, of
course, because the model does not violate the parity. We should
also note that the $F(R)$ gravity includes a scalar propagating
mode as in the Einstein frame action (\ref{FhGV7}). Even in the
Chern-Simons axion Einstein gravity, as clear from the action
(\ref{CS5}), there appears a propagating scalar mode. Both of the
scalar modes are massive and there is not so explicit difference.
However, regarding the gravitational waves, in the $f(R)$ gravity
case, one has a scalar component of the gravity waves, which is
though not present in the Chern-Simons axion Einstein case. The
only effect of the Chern-Simons term is to discriminate between
the two tensor modes of the gravitational wave, and in fact this
is the new feature that the Chern-Simons term induces, and it is a
challenge for the gravitational wave astronomy to find any parity
violating gravitational modes. A highly non-trivial task is to
investigate the $F(R)$ gravity extension of the axion Chern-Simons
gravity. We aim to study this case in a future work.

\section*{Acknowledgments}

This work is supported by MINECO (Spain), FIS2016-76363-P, and by
project 2017 SGR247 (AGAUR, Catalonia) (S.D.O). This work is also
supported by MEXT KAKENHI Grant-in-Aid for Scientific Research on
Innovative Areas ``Cosmic Acceleration'' No. 15H05890 (S.N.) and
the JSPS Grant-in-Aid for Scientific Research (C) No. 18K03615
(S.N.). This work is supported by the DAAD program
Hochschulpartnerschaften mit Griechenland 2016 (Projekt 57340132)
(V.K.O). V.K.O is indebted to Prof.~K.~Kokkotas for his
hospitality in the IAAT, University of T\"{u}bingen. The work of
A.P. is performed according to the Russian Government Program of
Competitive Growth of Kazan Federal University. The work of A.P.
was also supported by the Russian Foundation for Basic Research
Grant No 19-02-00496.

\section*{Appendix: Detailed Form of Einstein Tensor Components}

In this Appendix we present the explicit form of the $(t,t)$, $(i,j)$, and $(t,i)$
components of the Einstein tensor of Eq.~(\ref{CS15})
in the FRW background (\ref{FRWmetric}), which are,
\begin{align}
\label{tt}
G_{t t}=& -\dot H h_{tt} +H^2 \left[-3 h_{tt} +\frac{2}{a^2}\left( h_{xx}+h_{yy} +h_{zz} \right)
\right] \nn &
+\frac{H}{a^2} \left[ -\frac{\partial}{\partial t}\left( h_{xx}+h_{yy} +h_{zz} \right)
+2\left( \frac{\partial h_{tx}}{\partial x} + \frac{\partial h_{ty}}{\partial y}
+ \frac{\partial h_{tz}}{\partial z} \right) \right]
\nn &
+\frac{1}{a^4} \left[ \frac{1}{2} \left\{ \frac{\partial^2 }{\partial x^2}
\left( h_{yy} +h_{zz} \right) +\frac{\partial^2}{\partial y^2}\left( h_{xx} +h_{zz} \right)
+\frac{\partial^2}{\partial z^2}\left( h_{xx} +h_{yy} \right) \right\} \right.
\nn &
\left. -\frac{\partial^2 h_{xy}}{\partial x \partial y}
 -\frac{\partial^2 h_{xz}}{\partial x \partial z}
 -\frac{\partial^2 h_{yz}}{\partial y \partial z}\right]\, , \\
\label{xx}
G_{x x}=& \dot H \left[ a^2 h_{tt} -h_{yy} -h_{zz} +\frac{8 \dot U}{a}
\left( \frac{\partial h_{xz}}{\partial y}
-\frac{\partial h_{xy}}{\partial z}\right) \right] +H^2 \left( 3h_{tt} -h_{yy} -h_{zz} \right)
\nn &
+H \left[ a^2 \frac{\partial h_{tt}}{\partial t}
 - \frac{1}{2} \frac{\partial (h_{yy} +h_{zz})}{\partial t}
 -\frac{\partial h_{ty}}{\partial y} -\frac{\partial h_{tz}}{\partial z}
+\frac{8 \ddot U}{a} \left( \frac{\partial h_{xz}}{\partial y}
 -\frac{\partial h_{xy}}{\partial z}\right)
\right. \nn & \left.
+\frac{8 \dot U}{a} \left( \frac{\partial^2 h_{xz}}{\partial y \partial t}
 -\frac{\partial^2 h_{xy}}{\partial z \partial t}\right)  \right]
+ \frac{1}{2} \left( \frac{\partial^2 (h_{yy} +h_{zz})}{\partial t^2}
+\frac{\partial^2 h_{tt}}{\partial y^2} +\frac{\partial^2 h_{tt}}{\partial z^2}\right)
 -\frac{\partial^2 h_{ty}}{\partial y \partial t}
\nn &
 -\frac{\partial^2 h_{tz}}{\partial z \partial t}
+\frac{4 \ddot U}{a} \left( \frac{\partial^2 h_{tz}}{\partial y \partial x}
 -\frac{\partial^2 h_{ty}}{\partial z \partial x}
+\frac{\partial^2 h_{xy}}{\partial z \partial t}
 -\frac{\partial^2 h_{xz}}{\partial y \partial t}   \right)
+\frac{4 \dot U}{a} \left( \frac{\partial^3 h_{tz}}{\partial y \partial x \partial t}
\right. \nn & \left.
 -\frac{\partial^3 h_{ty}}{\partial z \partial x \partial t}
 -\frac{\partial^3 h_{xz}}{\partial y \partial t^2}
+\frac{\partial^3 h_{xy}}{\partial z \partial t^2} \right)
+\frac{1}{a^2} \left( \frac{\partial^2 h_{yz}}{\partial z \partial y}
 - \frac{1}{2} \frac{\partial^2 h_{yy}}{\partial z^2}
 - \frac{1}{2} \frac{\partial^2 h_{zz}}{\partial y^2} \right)
\nn &
+\frac{4 \dot U}{a^3} \left( \frac{\partial^3 h_{xz}}{ \partial y^3}
 -\frac{\partial^3 h_{xy}}{ \partial z^3}
 -\frac{\partial^3 h_{xy}}{\partial z \partial y^2}
+\frac{\partial^3 h_{xz}}{\partial y \partial z^2}
 -\frac{\partial^3 h_{yz}}{\partial x \partial y^2}
+\frac{\partial^3 h_{yz}}{\partial x \partial z^2} \right. \nn & \left.
+\frac{\partial^3 (h_{yy}-h_{zz})}{\partial x \partial y \partial z}\right)\, , \\
\label{yy}
G_{yy}=& \dot H \left[ a^2 h_{tt} -h_{xx} -h_{zz} -\frac{8 \dot U}{a}
\left( \frac{\partial h_{yz}}{\partial x}
 -\frac{\partial h_{xy}}{\partial z}\right) \right] +H^2 \left( 3h_{tt} -h_{xx} -h_{zz} \right)
\nn &
+H \left[ a^2 \frac{\partial h_{tt}}{\partial t}
 -\frac{1}{2} \frac{\partial (h_{xx} +h_{zz})}{\partial t}
 -\frac{\partial h_{tx}}{\partial x} -\frac{\partial h_{tz}}{\partial z}
 -\frac{8 \ddot U}{a} \left( \frac{\partial h_{yz}}{\partial x}
 -\frac{\partial h_{xy}}{\partial z}\right) \right. \nn & \left.
 -\frac{8 \dot U}{a} \left( \frac{\partial^2 h_{yz}}{\partial x \partial t}
 -\frac{\partial^2 h_{xy}}{\partial z \partial t}\right)  \right]
+ \frac{1}{2}\left( \frac{\partial^2 (h_{xx} +h_{zz})}{\partial t^2}
+\frac{\partial^2 h_{tt}}{\partial x^2} +\frac{\partial^2 h_{tt}}{\partial z^2} \right)
 -\frac{\partial^2 h_{tx}}{\partial x \partial t} \nn &
 -\frac{\partial^2 h_{tz}}{\partial z \partial t}
 -\frac{4 \ddot U}{a} \left( \frac{\partial^2 h_{tz}}{\partial y \partial x}
 -\frac{\partial^2 h_{tx}}{\partial z \partial y}
+\frac{\partial^2 h_{xy}}{\partial z \partial t}
 -\frac{\partial^2 h_{yz}}{\partial x \partial t} \right)
 -\frac{4 \dot U}{a} \left( \frac{\partial^3 h_{tz}}{\partial x \partial y \partial t}
\right. \nn & \left.
 -\frac{\partial^3 h_{tx}}{\partial z \partial y \partial t}
 -\frac{\partial^3 h_{yz}}{\partial x \partial t^2}
+\frac{\partial^3 h_{xy}}{\partial z \partial t^2} \right)
+\frac{1}{a^2} \left( \frac{\partial^2 h_{xz}}{\partial z \partial x}
 - \frac{1}{2} \frac{\partial^2 h_{xx}}{\partial z^2}
 - \frac{1}{2} \frac{\partial^2 h_{zz}}{\partial x^2} \right)
\nn &
 -\frac{4 \dot U}{a^3} \left( \frac{\partial^3 h_{yz}}{ \partial x^3}
 -\frac{\partial^3 h_{xy}}{ \partial z^3}
 -\frac{\partial^3 h_{xy}}{\partial z \partial x^2} +\frac{\partial^3 h_{yz}}{\partial x \partial z^2}
 -\frac{\partial^3 h_{xz}}{\partial y \partial x^2} +\frac{\partial^3 h_{xz}}{\partial y \partial z^2}
 \right. \nn & \left.
+\frac{\partial^3 (h_{xx}-h_{zz})}{\partial x \partial y \partial z}\right)\, , \\
\label{zz}
G_{zz}=& \dot H \left[ a^2 h_{tt} -h_{xx} -h_{yy}
+\frac{8 \dot U}{a}\left( \frac{\partial h_{yz}}{\partial x}
 -\frac{\partial h_{xz}}{\partial y}\right) \right] +H^2 \left( 3h_{tt}
 -h_{xx} -h_{yy} \right) \nn &
+H \left[ a^2 \frac{\partial h_{tt}}{\partial t}
 - \frac{1}{2} \frac{\partial (h_{xx} +h_{yy})}{\partial t}
 -\frac{\partial h_{tx}}{\partial x} -\frac{\partial h_{ty}}{\partial y}
 +\frac{8 \ddot U}{a} \left( \frac{\partial h_{yz}}{\partial x}
 -\frac{\partial h_{xz}}{\partial y}\right) \right. \nn & \left.
+\frac{8 \dot U}{a} \left( \frac{\partial^2 h_{yz}}{\partial x \partial t}
 -\frac{\partial^2 h_{xz}}{\partial y \partial t}\right)  \right]
+ \frac{1}{2} \left( \frac{\partial^2 (h_{xx} +h_{yy})}{\partial t^2}
+ \frac{\partial^2 h_{tt}}{\partial x^2} +\frac{\partial^2 h_{tt}}{\partial y^2} \right)
 -\frac{\partial^2 h_{tx}}{\partial x \partial t} \nn &
 -\frac{\partial^2 h_{ty}}{\partial y \partial t}
+\frac{4 \ddot U}{a} \left( \frac{\partial^2 h_{ty}}{\partial z \partial x}
 -\frac{\partial^2 h_{tx}}{\partial z \partial y}
+\frac{\partial^2 h_{xz}}{\partial y \partial t}
 -\frac{\partial^2 h_{yz}}{\partial x \partial t} \right)
+\frac{4 \dot U}{a} \left( \frac{\partial^3 h_{ty}}{\partial x \partial z \partial t}
\right. \nn & \left.
 -\frac{\partial^3 h_{tx}}{\partial y \partial z \partial t}
 -\frac{\partial^3 h_{yz}}{\partial x \partial t^2}
+\frac{\partial^3 h_{xz}}{\partial y \partial t^2} \right)
+\frac{1}{a^2} \left( \frac{\partial^2 h_{xy}}{\partial y \partial x}
 - \frac{1}{2} \frac{\partial^2 h_{xx}}{\partial y^2}
 - \frac{1}{2} \frac{\partial^2 h_{yy}}{\partial x^2} \right) \nn &
+\frac{4 \dot U}{a^3} \left( \frac{\partial^3 h_{yz}}{ \partial x^3}
 -\frac{\partial^3 h_{xz}}{ \partial y^3}
 -\frac{\partial^3 h_{xz}}{\partial y \partial x^2}
+\frac{\partial^3 h_{yz}}{\partial x \partial y^2}
 -\frac{\partial^3 h_{xy}}{\partial z \partial x^2}
+\frac{\partial^3 h_{xy}}{\partial z \partial y^2} \right. \nn & \left.
+\frac{\partial^3 (h_{xx}-h_{yy})}{\partial x \partial y \partial z}\right)\, , \\
\label{tx}
G_{t x}=& H \left[ \frac{\partial h_{tt}}{\partial x}
+\frac{1}{a^2}\left( \frac{\partial h_{xy}}{\partial y}
+\frac{\partial h_{xz}}{\partial z} -\frac{\partial h_{zz}}{\partial x}
 -\frac{\partial h_{yy}}{\partial x} \right)
\right. \nn & \left.
+\frac{4  \dot U}{a^3 } \left( \frac{\partial^2 h_{yz}}{\partial y^2}
 -\frac{\partial^2 h_{yz}}{\partial z^2}
+\frac{\partial^2 h_{xz}}{\partial x \partial y} -\frac{\partial^2 h_{xy}}{\partial x \partial z}
+\frac{\partial^2 h_{zz}}{\partial y \partial z} -\frac{\partial^2 h_{yy}}{\partial y \partial z}\right) \right]
\nn &
+\frac{1}{2 a^2}\left( \frac{\partial^2 h_{tx}}{\partial y^2} +\frac{\partial^2 h_{tx}}{\partial z^2}
+\frac{\partial^2 (h_{yy}+h_{zz})}{\partial x \partial t}
 -\frac{\partial^2 h_{xy}}{\partial t \partial y}
 -\frac{\partial^2 h_{xz}}{\partial t \partial z}
 -\frac{\partial^2 h_{ty}}{\partial x \partial y} -\frac{\partial^2 h_{tz}}{\partial x \partial z}
\right)
\nn &
+\frac{2 \dot U}{a^3 } \left( \frac{\partial^3 h_{tz}}{\partial y^3}
 -\frac{\partial^3 h_{ty}}{\partial z^3}
 -\frac{\partial^3 h_{xz}}{\partial t \partial x \partial y}
+\frac{\partial^3 h_{xy}}{\partial t \partial x \partial z}
+\frac{\partial^3 (h_{yy}-h_{zz})}{\partial t \partial y \partial z}
\right. \nn & \left.
 -\frac{\partial^3 h_{yz}}{\partial t \partial y^2} +\frac{\partial^3 h_{yz}}{\partial t \partial z^2}
+\frac{\partial^3 h_{tz}}{\partial y \partial x^2} -\frac{\partial^3 h_{ty}}{\partial z \partial x^2}
 -\frac{\partial^3 h_{ty}}{\partial z \partial y^2}
+\frac{\partial^3 h_{tz}}{\partial y \partial z^2}  \right)\, , \\
\label{ty}
G_{t y}=& H \left[ \frac{\partial h_{tt}}{\partial y}
+\frac{1}{a^2}\left( \frac{\partial h_{xy}}{\partial x}
+\frac{\partial h_{yz}}{\partial z} -\frac{\partial h_{zz}}{\partial y}
 -\frac{\partial h_{xx}}{\partial y} \right) \right. \nn & \left.
 -\frac{4  \dot U}{a^3 } \left( \frac{\partial^2 h_{xz}}{\partial x^2}
 -\frac{\partial^2 h_{xz}}{\partial z^2}
+\frac{\partial^2 h_{yz}}{\partial x \partial y}
 -\frac{\partial^2 h_{xy}}{\partial y \partial z}
+\frac{\partial^2 h_{zz}}{\partial x \partial z}
 -\frac{\partial^2 h_{xx}}{\partial x \partial z}\right) \right] \nn &
+\frac{1}{2 a^2}\left( \frac{\partial^2 h_{ty}}{\partial x^2} +\frac{\partial^2 h_{ty}}{\partial z^2}
+\frac{\partial^2 (h_{xx}+h_{zz})}{\partial y \partial t}
 -\frac{\partial^2 h_{xy}}{\partial t \partial x}
 -\frac{\partial^2 h_{yz}}{\partial t \partial z} -\frac{\partial^2 h_{tx}}{\partial x \partial y}
 -\frac{\partial^2 h_{tz}}{\partial y \partial z} \right) \nn &
 -\frac{2 \dot U}{a^3 } \left( \frac{\partial^3 h_{tz}}{\partial x^3}
 -\frac{\partial^3 h_{tx}}{\partial z^3}
 -\frac{\partial^3 h_{yz}}{\partial t \partial x \partial y}
+\frac{\partial^3 h_{xy}}{\partial t \partial y \partial z}
+\frac{\partial^3 (h_{xx}-h_{zz})}{\partial t \partial x \partial z} \right. \nn & \left.
 -\frac{\partial^3 h_{xz}}{\partial t \partial x^2}
+\frac{\partial^3 h_{xz}}{\partial t \partial z^2}
+\frac{\partial^3 h_{tz}}{\partial x \partial y^2} -\frac{\partial^3 h_{tx}}{\partial z \partial y^2}
 -\frac{\partial^3 h_{tx}}{\partial z \partial x^2}
+\frac{\partial^3 h_{tz}}{\partial x \partial z^2}  \right)\, , \\
\label{tz}
G_{t z}=& H \left[ \frac{\partial h_{tt}}{\partial z}
+\frac{1}{a^2}\left( \frac{\partial h_{zy}}{\partial y}
+\frac{\partial h_{xz}}{\partial x} -\frac{\partial h_{xx}}{\partial z}
 -\frac{\partial h_{yy}}{\partial z} \right)
\right. \nn & \left.
 -\frac{4  \dot U}{a^3 } \left( \frac{\partial^2 h_{yx}}{\partial y^2}
 -\frac{\partial^2 h_{yx}}{\partial x^2}
+\frac{\partial^2 h_{xz}}{\partial z \partial y} -\frac{\partial^2 h_{zy}}{\partial x \partial z}
+\frac{\partial^2 h_{xx}}{\partial y \partial x}
 -\frac{\partial^2 h_{yy}}{\partial y \partial x}\right) \right] \nn &
+\frac{1}{2 a^2}\left( \frac{\partial^2 h_{tz}}{\partial y^2} +\frac{\partial^2 h_{tz}}{\partial x^2}
+\frac{\partial^2 (h_{yy}+h_{xx})}{\partial z \partial t}
 -\frac{\partial^2 h_{zy}}{\partial t \partial y}
 -\frac{\partial^2 h_{xz}}{\partial t \partial x} -\frac{\partial^2 h_{ty}}{\partial z \partial y}
 -\frac{\partial^2 h_{tx}}{\partial x \partial z} \right) \nn &
 -\frac{2 \dot U}{a^3 } \left( \frac{\partial^3 h_{tx}}{\partial y^3}
 -\frac{\partial^3 h_{ty}}{\partial x^3}
 -\frac{\partial^3 h_{xz}}{\partial t \partial z \partial y}
+\frac{\partial^3 h_{zy}}{\partial t \partial x \partial z}
+\frac{\partial^3 (h_{yy}-h_{xx})}{\partial t \partial y \partial x} \right. \nn & \left.
 -\frac{\partial^3 h_{yx}}{\partial t \partial y^2}
+\frac{\partial^3 h_{yx}}{\partial t \partial x^2}
+\frac{\partial^3 h_{tx}}{\partial y \partial z^2} -\frac{\partial^3 h_{ty}}{\partial x \partial z^2}
 -\frac{\partial^3 h_{ty}}{\partial x \partial y^2}
+\frac{\partial^3 h_{tx}}{\partial y \partial x^2}  \right)\, , \\
\label{xy}
G_{xy}=& \dot H \left[ h_{xy} +\frac{4 \dot U}{a} \left( \frac{\partial (h_{xx}-h_{yy})}{\partial z}
+\frac{\partial h_{yz}}{\partial y} -\frac{\partial h_{xz}}{\partial x} \right) \right]
+H^2 h_{xy} \nn
& +H \left[ \frac{1}{2} \left( \frac{\partial h_{tx}}{\partial y} +\frac{\partial h_{ty}}{\partial x}
+\frac{\partial h_{xy}}{\partial t} \right)
+\frac{4 \ddot U}{a}\left( \frac{\partial (h_{xx}-h_{yy})}{\partial z}
+\frac{\partial h_{yz}}{\partial y} -\frac{\partial h_{xz}}{\partial x} \right)
\right. \nn & \left.
+\frac{4 \dot U}{a}\left( \frac{\partial^2 (h_{xx}-h_{yy})}{\partial z \partial t}
+\frac{\partial^2 h_{yz}}{\partial y \partial t}
 -\frac{\partial^2 h_{xz}}{\partial x \partial t} \right)\right]
+ \frac{1}{2} \left( \frac{\partial^2 h_{tx}}{\partial y \partial t}
+\frac{\partial^2 h_{ty}}{\partial x \partial t}
 -\frac{\partial^2 h_{tt}}{\partial y \partial x} -\frac{\partial^2 h_{xy}}{\partial t^2} \right)
\nn &
+\frac{2 \ddot U}{a}\left( \frac{\partial^2 (h_{yy}-h_{xx})}{\partial z \partial t}
 -\frac{\partial^2 h_{tz}}{\partial x^2} +\frac{\partial^2 h_{tz}}{\partial y^2}
+\frac{\partial^2 h_{tx}}{\partial z \partial x} -\frac{\partial^2 h_{ty}}{\partial z \partial y}
+\frac{\partial^2 h_{xz}}{\partial x \partial t}
 -\frac{\partial^2 h_{yz}}{\partial y \partial t} \right) \nn &
+\frac{2 \dot U}{a}\left( \frac{\partial^3 (h_{yy} - h_{xx})}{\partial z \partial t^2}
+\frac{\partial^3 h_{tx}}{\partial z \partial x \partial t}
 -\frac{\partial^3 h_{ty}}{\partial z \partial y \partial t}
 -\frac{\partial^3 h_{tz}}{\partial x^2 \partial t} +\frac{\partial^3 h_{tz}}{\partial y^2 \partial t}
+\frac{\partial^3 h_{xz}}{\partial x \partial t^2}
 -\frac{\partial^3 h_{yz}}{\partial y \partial t^2} \right) \nn &
+\frac{1}{2 a^2} \left( \frac{\partial^2 h_{xy}}{\partial z^2}
+\frac{\partial^2 h_{zz}}{\partial y \partial x}
 -\frac{\partial^2 h_{xz}}{\partial z \partial y}
 -\frac{\partial^2 h_{yz}}{\partial z \partial x} \right)
+\frac{2 \dot U}{a^3}\left( \frac{\partial^3 (h_{xx}-h_{yy})}{\partial z^3}
+\frac{\partial^3 (h_{zz}-h_{yy})}{\partial z \partial x^2} \right. \nn & \left.
+\frac{\partial^3 (h_{xx}-h_{zz})}{\partial z \partial y^2}
+2\frac{\partial^3 h_{yz}}{\partial y \partial x^2}
 -2\frac{\partial^3 h_{xz}}{\partial x \partial y^2}
+2\frac{\partial^3 h_{yz}}{\partial y \partial z^2}
 -2\frac{\partial^3 h_{xz}}{\partial x \partial z^2}  \right)\, , \\
\label{xz}
G_{xz}=& \dot H \left[ h_{xz} -\frac{4 \dot U}{a} \left( \frac{\partial (h_{xx}-h_{zz})}{\partial y}
+\frac{\partial h_{yz}}{\partial z} -\frac{\partial h_{xy}}{\partial x} \right) \right]
+H^2 h_{xz} \nn
& +H \left[ \frac{1}{2} \left( \frac{\partial h_{tx}}{\partial z} +\frac{\partial h_{tz}}{\partial x}
+\frac{\partial h_{xz}}{\partial t} \right)
 -\frac{4 \ddot U}{a}\left( \frac{\partial (h_{xx}-h_{zz})}{\partial y}
+\frac{\partial h_{yz}}{\partial z} -\frac{\partial h_{xy}}{\partial x} \right)
\right. \nn & \left.
 -\frac{4 \dot U}{a}\left( \frac{\partial^2 (h_{xx}-h_{zz})}{\partial y \partial t}
+\frac{\partial^2 h_{yz}}{\partial z \partial t}
 -\frac{\partial^2 h_{xy}}{\partial x \partial t} \right)\right]
+ \frac{1}{2} \left( \frac{\partial^2 h_{tx}}{
\partial z \partial t} +\frac{\partial^2 h_{tz}}{\partial x \partial t}
 -\frac{\partial^2 h_{tt}}{\partial z \partial x} -\frac{\partial^2 h_{xz}}{\partial t^2} \right)
\nn &
 -\frac{2 \ddot U}{a}\left( \frac{\partial^2 (h_{zz}-h_{xx})}{\partial y \partial t}
 -\frac{\partial^2 h_{ty}}{\partial x^2} +\frac{\partial^2 h_{ty}}{\partial z^2}
+\frac{\partial^2 h_{tx}}{\partial y \partial x} -\frac{\partial^2 h_{tz}}{\partial z \partial y}
+\frac{\partial^2 h_{xy}}{\partial x \partial t}
 -\frac{\partial^2 h_{yz}}{\partial z \partial t} \right) \nn &
 -\frac{2 \dot U}{a}\left( \frac{\partial^3 (h_{zz} -h_{xx})}{\partial y \partial t^2}
+\frac{\partial^3 h_{tx}}{\partial y \partial x \partial t}
 -\frac{\partial^3 h_{tz}}{\partial z \partial y \partial t}
 -\frac{\partial^3 h_{ty}}{\partial x^2 \partial t} +\frac{\partial^3 h_{ty}}{\partial z^2 \partial t}
+\frac{\partial^3 h_{xy}}{\partial x \partial t^2}
 -\frac{\partial^3 h_{yz}}{\partial z \partial t^2} \right) \nn &
+\frac{1}{2 a^2} \left( \frac{\partial^2 h_{xz}}{\partial y^2}
+\frac{\partial^2 h_{yy}}{\partial z \partial x}
 -\frac{\partial^2 h_{xy}}{\partial z \partial y}
 -\frac{\partial^2 h_{yz}}{\partial y \partial x} \right)
 -\frac{2 \dot U}{a^3}\left( \frac{\partial^3 (h_{xx}-h_{zz})}{\partial y^3}
+\frac{\partial^3 (h_{yy}-h_{zz})}{\partial y \partial x^2}
\right. \nn & \left.
+\frac{\partial^3 (h_{xx}-h_{yy})}{\partial y \partial z^2}
+2\frac{\partial^3 h_{yz}}{\partial z \partial x^2}
 -2\frac{\partial^3 h_{xy}}{\partial x \partial z^2}
+2\frac{\partial^3 h_{yz}}{\partial z \partial y^2}
 -2\frac{\partial^3 h_{xy}}{\partial x \partial y^2}  \right)\, , \\
\label{yz}
G_{yz}=& \dot H \left[ h_{yz} +\frac{4 \dot U}{a} \left( \frac{\partial (h_{yy}-h_{zz})}{\partial x}
+\frac{\partial h_{xz}}{\partial z} -\frac{\partial h_{xy}}{\partial y} \right) \right]
+H^2 h_{yz} \nn
& +H \left[ \frac{1}{2} \left( \frac{\partial h_{ty}}{\partial z} +\frac{\partial h_{tz}}{\partial y}
+\frac{\partial h_{yz}}{\partial t} \right)
+\frac{4 \ddot U}{a}\left( \frac{\partial (h_{yy}-h_{zz})}{\partial x}
+\frac{\partial h_{xz}}{\partial z} -\frac{\partial h_{xy}}{\partial y} \right)
\right. \nn & \left.
+\frac{4 \dot U}{a}\left( \frac{\partial^2 (h_{yy}-h_{zz})}{\partial x \partial t}
+\frac{\partial^2 h_{xz}}{\partial z \partial t}
 -\frac{\partial^2 h_{xy}}{\partial y \partial t} \right)\right]
+ \frac{1}{2} \left( \frac{\partial^2 h_{ty}}{\partial z \partial t}
+\frac{\partial^2 h_{tz}}{\partial y \partial t}
 -\frac{\partial^2 h_{tt}}{\partial z \partial y} -\frac{\partial^2 h_{yz}}{\partial t^2} \right)
\nn &
+\frac{2 \ddot U}{a}\left( \frac{\partial^2 (h_{zz}-h_{yy})}{\partial x \partial t}
-\frac{\partial^2 h_{tx}}{\partial y^2} +\frac{\partial^2 h_{tx}}{\partial z^2}
+\frac{\partial^2 h_{ty}}{\partial y \partial x} -\frac{\partial^2 h_{tz}}{\partial z \partial x}
+\frac{\partial^2 h_{xy}}{\partial y \partial t}
 -\frac{\partial^2 h_{xz}}{\partial z \partial t} \right) \nn &
+\frac{2 \dot U}{a}\left( \frac{\partial^3 (h_{zz}-h_{yy})}{\partial x \partial t^2}
+\frac{\partial^3 h_{ty}}{\partial y \partial x \partial t}
 -\frac{\partial^3 h_{tz}}{\partial z \partial x \partial t}
 -\frac{\partial^3 h_{tx}}{\partial y^2 \partial t} +\frac{\partial^3 h_{tx}}{\partial z^2 \partial t}
+\frac{\partial^3 h_{xy}}{\partial y \partial t^2}
 -\frac{\partial^3 h_{xz}}{\partial z \partial t^2} \right) \nn &
+\frac{1}{2 a^2} \left( \frac{\partial^2 h_{yz}}{\partial x^2}
+\frac{\partial^2 h_{xx}}{\partial z \partial y}
 -\frac{\partial^2 h_{xy}}{\partial z \partial x}
 -\frac{\partial^2 h_{xz}}{\partial y \partial x} \right)
+\frac{2 \dot U}{a^3}\left( \frac{\partial^3 (h_{yy}-h_{zz})}{\partial x^3}
+\frac{\partial^3 (h_{xx}-h_{zz})}{\partial x \partial y^2}
\right. \nn & \left.
+\frac{\partial^3 (h_{yy}-h_{xx})}{\partial x \partial z^2}
+2\frac{\partial^3 h_{xz}}{\partial z \partial y^2}
 -2\frac{\partial^3 h_{xy}}{\partial y \partial z^2}
+2\frac{\partial^3 h_{xz}}{\partial z \partial x^2}
 -2\frac{\partial^3 h_{xy}}{\partial y \partial x^2}  \right)\, .
\end{align}

\end{document}